\documentstyle[preprint,aps]{revtex}

\newcommand{\be}{\begin{eqnarray}}
\newcommand{\ee}{\end{eqnarray}}

\newcommand{\nsz}{\textstyle}

\begin{document}
\preprint{SUNY-NTG-95-22}
\draft
\title{ The Interacting Instanton Liquid \\
        in QCD at zero and finite Temperature}

\author{T.~Sch\"afer and E.V.~Shuryak }

\address{Department of Physics, State University of New York at Stony
         Brook,\\ Stony Brook, New York 11794, USA}

\date{\today}
\maketitle

\begin{abstract}
   In this paper we study the statistical mechanics of the instanton
liquid in QCD. After introducing the partition function as well as
the gauge field and quark induced interactions between instantons
we describe a method to calculate the free energy of the instanton
system. We use this method to determine the equilibrium density and
the equation of state from numerical simulations of the instanton
ensemble in QCD for various numbers of flavors. We find that there
is a critical number of flavors above which chiral symmetry is
restored in the groundstate. In the physical case of two light
and one intermediate mass flavor the system undergoes a chiral phase
transition at $T\simeq 140$ MeV. We show that the mechanism for this
transition is a rearrangement of the instanton liquid, going from a
disordered, random, phase at low temperatures to a strongly correlated,
molecular, phase at high temperature. We also study the behavior of
mesonic susceptibilities near the phase transition.
\end{abstract}

\section{Introduction}

   Understanding the vacuum structure of gauge theories like QCD
is one of the main problems in quantum field theory today. It also
provides the theoretical foundation for hadronic models and hadronic
phenomenology from the underlying field theory of the strong interaction,
Quantum Chromodynamics. There are a number of indications that instantons,
classical tunneling trajectories in imaginary (euclidean) time are an
important ingredient of the QCD vacuum.

   Soon after the discovery of instantons 20 years ago \cite{BPST_75},
it became clear that instantons may provide at least a qualitative
understanding of many features of the QCD vacuum. Instantons solve
the $U(1)_A$ problem \cite{tHo_76}, they give a mechanism for chiral
symmetry breaking \cite{CDG_78}, contribute to the gluon condensates
and lead to a non-perturbative vacuum energy \cite{Shu_78,SVZ_79}.

   The development of a quantitative theory based on these ideas took much
longer. The instanton liquid model was originally suggested by Shuryak
in 1982 \cite{Shu_82}, guided mainly by phenomenological considerations.
Later Diakonov and Petrov developed an analytical approach based on the
variational method \cite{DP_84} (see also \cite{IM_81}). The first numerical
simulations of the ``instanton liquid" were reported in \cite{Shu_88,Shu_89}.
During the past two years we have shown \cite{SV_93,SS_94,SS_95} that the
``random instanton liquid model" (RILM) provides a successful description
of a large number of hadronic correlation functions, including mesons and
baryons made of light quarks, heavy-light systems and glueballs. These
correlators not only give reasonable values for the corresponding resonance
masses and coupling constants, but they also compare well with
point-to-point correlation functions extracted from phenomenology
\cite{Shu_93} or measured on the lattice \cite{CGHN_93}.

    Following these developments, several recent lattice simulations
have focused on the role of instantons in the QCD vacuum, both
at zero and finite temperature. Using a method called ''cooling", one
can relax any given gauge field configuration to the closest classical
component of the QCD vacuum. The resulting configurations were known to be of
multi-instanton type \cite{Ber_81}, but the more recent work by Chu et
al.~\cite{CGHN_94} has provided quantitative measurements of the parameters
of the instanton liquid, as well as detailed studies of the dynamical
effects of instantons. These authors conclude that the instanton density in
the quenched theory (without dynamical fermions) at zero temperature is
$n\simeq (1.3$-$1.6)\,{\rm fm}^{-4}$ while the average size is about
$\rho\simeq 0.35$ fm. These numbers confirm the key parameters $n=1\,
{\rm fm}^{-4}$ and $\rho=1/3$ fm of the instanton liquid model mentioned
above. In addition to that, Chu et al.~studied correlation functions
in the cooled configurations, finding that they hardly change from the
original, fully quantum configurations. This would imply that instanton
effects dominate over perturbative and confinement forces in determining
the structure of low-lying hadronic states. A lattice measurement of the
instanton size distribution for pure $SU(2)$ gauge theory was performed in
\cite{MS_95}, where also an attempt was made to study correlations
between instantons. The size distribution was compared to the predictions
of the interacting instanton liquid model in \cite{Shu_95}.

   In this paper we want to report a detailed study of the statistical
mechanics of the interacting instanton liquid (IILM), both at zero and
finite temperature. The purpose of this study is twofold. First, we
want to give a much more consistent treatment of the model at zero
temperature. Second, we explore the model at non-zero temperature. In
this paper, we will limit ourselves to the analysis of bulk properties,
like the energy density, the chiral condensate and mesonic susceptibilities.
A detailed study of hadronic correlation functions will be presented in a
forthcoming publication \cite{SS_95c}.

   At zero temperature, we want to study the importance of instanton
interactions and determine those features of the interacting ensemble
that differ from the simplest, random ensemble (RILM). We also want
to determine the relation between the instanton interaction and global
parameters (like the density and average size) of the instanton
liquid. Furthermore, we want to quantify the role of instanton
interactions in producing correlations among the instantons. This
is of interest, since despite the success of the random model in the
description of most hadronic correlation functions \cite{SV_93}, it
fails in channels where the single instanton interaction is strongly
repulsive (like the $\eta'$ and $\delta$ meson channels). As shown in
\cite{SV_95,SS_95,Zah_94}, the correlations among instantons caused by
their classical and fermion induced interactions lead to a correct
description of topological charge screening and the $\eta'$ channel.

   The role of correlations among the instantons is particularly
important with regard to the nature of the chiral phase transition.
Originally, it was believed \cite{Shu_82,DM_88,NVZ_89b} that chiral
symmetry is restored due to the rapid disappearance of instantons
at high temperature \cite{PY_80}. Ilgenfritz and Shuryak realized
\cite{IS_89} that instantons can be present even above the chiral
phase transition, as strongly correlated instanton-antiinstanton
molecules (see also \cite{KY_91}). In this case, the transition is
determined by the phase equilibrium between the low temperature
``liquid" and high temperature ``molecular" phase. Shuryak and
Velkovsky later argued that instanton suppression is essentially a
plasma effect and should not be present below the phase transition
\cite{SV_94}. This means that the transition is driven by the formation
of molecules \cite{IS_94,SSV_95} rather than by the suppression of
individual instantons. There is some support for this scenario from
lattice simulations performed by Chu and Schramm \cite{CS_95}. Extending
the cooling method to finite temperature, they find that the instanton
density is essentially independent of temperature below the phase transition,
while it is exponentially suppressed above the transition temperature.
First evidence for the presence of molecules near $T_c$ was reported
in \cite{IMM_95}.

    So far, the transition has only been studied using the schematic
''cocktail" model introduced by Ilgenfritz and Shuryak \cite{IS_94}.
In this model the instanton liquid consists of two components, a random
and a molecular one. The free energy is determined separately for the
two components and their concentrations is then determined by minimizing
the total free energy. The chiral phase transition occurs when the
concentration of random instantons is zero. This approach predicts
the presence of a substantial number of instantons even if $T>T_c$,
causing new non-perturbative effects in the plasma phase. One such
effect, studied in \cite{SSV_95}, is the ``molecule-induced"
effective interaction between quarks, leading to a spectrum of
spacelike screening masses consistent with lattice data. Another
possible consequence, the survival of certain hadronic modes
above the phase transition was studied in \cite{SS_95b}.

    In the present work we significantly improve on the schematic
model used in these works, by doing a complete calculation in the
interacting ensemble. In this way, many approximations are relaxed
and all possible correlations among instantons (not just polarized
instanton-antiinstanton pairs) are included. The paper is organized
as follows. In section 2 we introduce the partition function of the
instanton liquid and specify the gauge field and fermion induced
interactions between instantons. Detailed parameterizations of these
interactions can be found in the appendices. In section 3 we describe
the method which is used to calculate the partition function.
In section 4 we use this method to study the instanton ensemble at
zero temperature. In sections 5 and 6 we generalize the method
to finite temperature and study the nature of the chiral phase
transition in the instanton liquid. In section 7, we study some
phenomena associated with the transition, in particular the Dirac
spectrum and the mesonic susceptibilities.

\section{The partition function of the instanton liquid}

    The euclidean partition function of QCD is given by
\be
\label{Z_QCD}
  Z = \int DA_\mu\, \exp(-S[A_\mu])
          \prod_f^{N_f}\det(\hat D+m_f)
\ee
where the gauge field action is given by $S[A_\mu]=\frac{1}{4}
\int d^4x\, {\rm Tr}(F_{\mu\nu}F_{\mu\nu})$ and the Dirac operator
is defined by $\hat D\psi=\gamma_\mu(\partial_\mu -iA_\mu)\psi$.

    The main assumption underlying the instanton model is that the
full partition function can be approximated by a partition sum in
which the relevant gauge configurations are superpositions of instantons
and antiinstantons. In this partition function the integration
extends over the collective coordinates associated with $N_+$
instantons and $N_-$ instantons
\be
\label{Z}
Z =  {1 \over N_+ ! N_- !}\int
    \prod_i^{N_+ + N_-} [d\Omega_i\; d(\rho_i) ]
    \exp(-S_{int})\prod_f^{N_f} \det(\hat D+m_f) \, .
\ee
Here $d\Omega_i=dU_i\, d^4z_i\,d\rho_i$ is the measure in the space of
collective coordinates, color orientation, position and size, associated
with single instantons. For the gauge group $SU(3)$, there are 12
collective coordinates per instanton. Fluctuations around individual
instantons are included in gaussian approximation. This gives the
semi-classical instanton amplitude, originally calculated by 't Hooft
\cite{tHo_76}. To two-loop accuracy, it reads
\be
\label{idens}
d(\rho) &=& C_{N_c} \rho^{-5} \beta_1 (\rho)^{2N_c}
            \exp\left(-\beta_2 (\rho)+(2N_c-\frac{b'}{2b})
            \frac{b'}{2b}\frac{1}{\beta_1(\rho)}
            \log (\beta_1(\rho))
            \right) \, ,\\
  & & C_{N_c} = \frac{ 4.6\exp(-1.86 N_c) }
                 {\pi^2 (N_c-1)! (N_c-2)! }\,  ,
\ee
where $\beta_1(\rho)$ and $\beta_2(\rho)$ are the one and two loop
beta functions
\be
\beta_1(\rho) = -b\log(\rho\Lambda), \hspace{1cm}
\beta_2(\rho) = \beta_1(\rho) + \frac{b'}{2b}
   \log (\frac{2}{b}\beta_1(\rho)) ,
\ee
with the one and two loop coefficients
\be
   b = \frac{11}{3}N_c -\frac{2}{3}N_f , \hspace{1cm}
   b'= \frac{34}{3}N_c^2-\frac{13}{3}N_c N_f +\frac{N_f}{N_c}.
\ee
The coefficient $C_{N_c}$ was calculated in a Pauli-Vilars renormalization
scheme, and $\Lambda$ is the corresponding scale parameter.
The classical action $S_0 = \frac{8\pi^2}{g^2}$ is included in the
semi-classical amplitude (\ref{idens}). The classical interaction
between instantons is denoted by $S_{int}$. We approximate this
interaction by a pure two-body interaction $S_{int}=\frac{1}{2}\sum_{I\neq J}
S_{int}(\Omega_{IJ})$ which only depends on the relative coordinates
of the two instantons. The importance of genuine three body effects
in the classical interaction between instantons was studied in
\cite{Shu_88}, with the conclusion that this contribution is
negligible as long as the density is not extremely large.

    The two-body interaction $S_{int}(\Omega_{IJ}) = S[A_\mu(
\Omega_{IJ})]-2S_0$ is calculated classically, by inserting the
two-instanton gauge potential $A_\mu(\Omega_{IJ})$ into the action.
There are no exact instanton-antiinstanton solutions to the classical
Yang-Mills equations, so in practice one has to use an ansatz for
the gauge potential. The resulting interaction will then depend on the
details of the trial function. Various trial functions have been used in the
literature: (i) the sum ansatz \cite{DP_84}, (ii) the ratio ansatz
\cite{Shu_88}, (iii) the conformaly invariant Yung ansatz \cite{Yun_88}
and (iv) the exact streamline solution \cite{Ver_91}. The latter is
characterized by the fact that the action is minimized in all directions
except along the collective coordinate describing the separation between
the two instantons. In this sense, the streamline solution is the optimal
classical instanton-antiinstanton configuration.

   In order to discuss the properties of the classical interaction
between instantons, let us introduce the four vector $u_\mu =
\frac{1}{2i}{\rm tr}(U_I^+ U_A \tau_\mu^{+})$ where $U_{I,A}$ are
the orientation matrices of the instanton and antiinstanton and
$\tau_\mu^+$ is the $2\times 2$ matrix $(\vec\tau,-i)$. For the gauge
group $SU(2)$, $u_\mu$ is a real unit vector whereas for $SU(3)$
it is a complex vector with $|u|^2\leq 1$. In any case, we can define
an angle $\theta$ by
\be
   \cos\theta = \frac{|u\cdot\hat R|}{|u|},
\ee
where $R=z_I-z_A$ is the vector connecting the centers of the two
instantons.  For all the trial functions mentioned above, the large
distance part of the instanton-antiinstanton interaction is given by
\be
\label{dipole}
  S_{int} = \beta_1(\bar\rho)\frac{4\rho_1^2\rho_2^2}{R^4}|u|^2
                  (1-4\cos^2\theta),
\ee
which is the dipole-dipole interaction originally discussed by Callan,
Dashen and Gross \cite{CDG_78}. The interaction is given in units of
the single instanton action $\beta_1(\rho)$. The argument of the
beta function is not exactly determined without a calculation of
the fluctuations around the two-instanton configuration.
In practice, we take the argument to be the geometric mean $\bar\rho=
\sqrt{\rho_I\rho_A}$ of the two instanton radii. The interaction is
attractive for the relative orientation $\cos\theta=1$, but vanishes after
averaging over all angles $\theta$. The short distance behavior depends
on the ansatz chosen. In the sum ansatz, there is a substantial
repulsive core at distances $R<\sqrt{6}\rho$ \cite{DP_84}, but the
amount of repulsion at short distances becomes significantly weaker
using the more refined trial functions. We will discuss this question
in more detail in the next section. A parameterization of the interaction
in the ratio and streamline ansaetze is given in appendix A. In
fig.1a, we show the ratio and streamline interaction for the most
attractive and repulsive orientations. One clearly observes that the
interactions are similar at large distance, but differ significantly at
short distance. In particular, the streamline interaction has no
repulsion at all for the most attractive orientation. The interaction
smoothly approaches $S_{int}=-2S_0$ at short distance, corresponding
to the annihilation of the instanton-antiinstanton pair. Fig.1a also
shows the effect of a phenomenological core in the streamline interaction.
The reasoning behind this modification will be discussed in more detail in
the next section.

   For all the trial functions except for the simple sum ansatz, the
instanton-instanton interaction is much weaker than the
instanton-antiinstanton one. In fact, in the streamline ansatz, the
instanton-instanton interaction vanishes. This is a reflection
of the fact that there is an exact two-instanton solution (with $S=2S_0$)
for arbitrary values of the relative coordinates.

   The fermionic determinant induces a very nonlocal interaction
among the instantons. Evaluating this determinant exactly in the
instanton ensemble still constitutes a formidable problem. In practice
we factorize the determinant into a low and a high momentum part
\cite{DP_86}
\be
\label{det}
  \det(\hat D+m_f) = \left( \prod_{i}^{N_++N_-}\hspace{-0.3cm}
  1.34 \rho_i\right) \;\det(T+m_f),
\ee
where the first factor, the high momentum part, is the product of
contributions from individual instantons calculated in gaussian
approximation, whereas the low momentum part associated with the
fermionic zero modes of individual instantons is calculated exactly.

   This means that the instanton induced 't Hooft interaction between
quarks \cite{tHo_76,SVZ_80} is included to all orders. The low momentum
part of the spectrum of the Dirac operator is also of special significance
in connection with the structure of chiral symmetry breaking. $T_{IA}$
is the $N_+\times N_-$ matrix of overlap matrix elements
\be
\label{overl}
 T_{IA} &=& \int d^4x\;  \phi_{A\, 0}^\dagger (x-z_A) i\hat D_x
       \phi_{I\, 0}(x-z_I),
\ee
where $\phi_{I,A\, 0}$ are the fermionic zero mode wave functions
of the instanton and antiinstanton. Due to the chirality of the
zero modes, the fermionic overlap matrix elements between instantons
with the same topological charge vanishes. In the following, we will only
consider quadratic matrices with $N_+=N_-$. Ensembles with $N_+\neq
N_-$ correspond to systems with net topological charge. In the
thermodynamic limit, the distribution of winding numbers is sharply
peaked around zero \cite{LS_92}, and imposing an additional condition
corresponding to a finite winding number is not expected to affect
our results. Fluctuations of the topological charge can be studied
by considering appropriately chosen subsytems \cite{SV_95}.

    The general structure of the overlap matrix elements is given
by $T_{IA}= (u\cdot R) f(R)$.  This means that, like the gauge
field induced interaction, the fermionic overlap is maximal when
the relative instanton-antiinstanton orientation is given by
$\cos\theta=1$. Like the gauge field induced interaction
between instantons, the fermionic overlap matrix elements depend on
the ansatz for the two-instanton gauge potential. In this case, however,
the dependence on the trial function is much less pronounced. For the sum
ansatz, one can use the equations of motion and replace the covariant
derivative in (\ref{overl}) by an ordinary one. The result can be
parameterized by
\be
\label{sum_overl}
 T_{IA} = i(u\cdot R)  \frac{1}{\rho_I\rho_A}
    \frac{4.0}{(2.0+R^2/\rho_I\rho_A)^2},
\ee
which is exact at large distances. The streamline ansatz gives the
same large distance behavior, but somewhat different results at
small and intermediate separations. We give a parameterization of
the streamline matrix elements in appendix B.

\section{The free energy of the instanton ensemble}

   In this section, we describe a method to evaluate the partition
function of the instanton liquid. Using this method, we can calculate
the free energy numerically as a function of the density of instantons,
and determine the equilibrium density from the condition that the free
energy is minimal.

   The problem in determining the free energy is connected with the fact
that the complicated statistical mechanics associated with the partition
function (\ref{Z}) can in general only be dealt with by performing
Monte Carlo simulations \cite{Shu_88,Shu_89,SV_90}. These simulations
are ideally suited for the calculation of various expectation values,
but do not give a direct determination of the partition function,
which provides the overall normalization. Previous Monte Carlo calculations
have therefore been restricted to simulations of the ensemble at a fixed
density of instantons, which was determined from phenomenological
considerations (typically $1\,{\rm fm}^{-4}$). Here, we want to go
beyond this approximation and minimize the free energy. A method
to calculate the partition sum, which is well known in statistical
mechanics, quantum mechanics (see, e.g. \cite{SZ_84}) and lattice
gauge theory \cite{EFK*90}, is ''adiabatic switching". For this
purpose one writes the effective action as
\be
S_{eff} = S_{0}+\alpha S_{1},
\ee
which interpolates between a solvable action $S_0$ and the full
action $S_0+S_1$. If the partition function for the system
governed by the action $S_0$ is known, the full partition function
can be determined from
\be
\label{int_coup}
  \log Z(\alpha\! =\! 1) &=& \log Z(\alpha\! =\! 0)
  - \int_0^1 d\alpha'\, \langle 0| S_1 |0\rangle_{\alpha'},
\ee
where the expectation value $\langle 0|.|0\rangle_{\alpha}$ depends on
the coupling constant $\alpha$. In our case, the effective action is
given by
\be
\label{S_eff}
   S_{eff} = -\sum_{i=1}^{N_++N_-}\log (d(\rho_i)) + S_{int}
               +{\rm tr}\log (\hat D+m_f)\, .
\ee
The obvious choice for decomposing the effective action of the
instanton liquid would be to identify the logarithm of the single
instanton distribution with the free action, $S_0=\sum_i\log(d(\rho_i))$.
This procedure, however, does not work since the instanton distribution
behaves like $d(\rho) \sim \rho^{(b-5)}$, so that the $\rho$ integration
in the free partition function would not be convergent. This is the
famous infrared problem which plagues the dilute instanton gas approximation
\cite{CDG_78}. As explained in more detail in the next section, the instanton
liquid is stabilized by the the repulsive core in the gauge field interaction
once the full interaction is taken into account. We therefore consider
the following decomposition
\be
\label{S_var}
   S_{eff} = \sum_{i=1}^{N_++N_-}\left(- \log (d(\rho_i)) +
           (1-\alpha)\nu\frac{\rho_i^2}{\,\overline{\rho^2}\,}\right)
           + \alpha \left( S_{int}
               +{\rm tr}\log (\hat D+m_f) \right),
\ee
where $\nu=(b-4)/2$ and $\overline{\rho^2}$ is the average size squared
of the instantons with the full interaction included. The term
proportional to $(1-\alpha)$ serves to regularize the $\rho$
integration for $\alpha=0$. It disappears for $\alpha=1$, where
the original action is recovered. The specific form of this
term is irrelevant, our choice here is motivated by the fact
that (at least for the one-loop measure $d(\rho)$) $S_{eff}
(\alpha$=$0)$ yields a single instanton distribution with the correct
average size $\overline{\rho^2}$. This means that the single instanton
distribution generated by $S_{eff}(\alpha$=$0)$ is a variational
ansatz for the full one-instanton distribution.

   The partition function corresponding to the variational single
instanton distribution is given by
\be
\label{Z_free}
   Z_0 = \frac{1}{N_+!\, N_-!} (V\mu_0)^{N_++N_-}, \hspace{1cm}
         \mu_0 = \int_0^\infty  d\rho \, d(\rho)
           \exp(-\nu\frac{\rho^2}{\,\overline{\rho^2}\,} ) ,
\ee
where $\mu_0$ is the normalization of the one-body distribution.
The $\rho$ integration in $\mu_0$ is regularized by the second term
in (\ref{S_var}). The full partition function obtained from integrating
over the coupling  $\alpha$ is
\be
\label{int_coup2}
  \log Z &=& \log (Z_0)
  + N \int_0^1 d\alpha'\,  \langle 0|
      \nu\frac{\rho^2}{\,\overline{\rho^2}\,} - \frac{1}{N}
      \left( S_{int}+{\rm tr}\log(\hat D+m_f) \right)
      |0\rangle_{\alpha'},
\ee
where $N=N_++N_-$. The free energy density is finally given by $F=-1/V\cdot
\log Z$ where $V$ is the four-volume of the system. The pressure and the
energy density are related to $F$ by
\be
 p=-F, \hspace{1cm}
 \epsilon = T \frac{dp}{dT}-p .
\ee
At zero temperature we have $\epsilon=-p=F$ and the free energy determines
the shift of the QCD ground state relative to the perturbative vacuum.
Such a shift is certainly present in our case, since tunneling lowers
the ground state energy.

    The full partition function can be compared to the variational
ansatz introduced in \cite{DP_84} and employed in many works on
the subject \cite{Shu_85,DM_88,IS_89,NVZ_89b,Ver_91,IS_94}. For
simplicity we restrict the discussion at this point to pure gauge
theory, i.e. neglect the fermionic determinant. Since the variational
ansatz ignores any correlations between instantons, only the color and
spatial average of the interaction enters
\be
 \int d^4R dU S_{int}(R,U,\rho_1,\rho_2)
    = \kappa^2 \frac{N_c}{N_c^2-1}\rho_1^2 \rho_2^2 ,
\ee
where $S_{int}(R,U,\rho_1,\rho_2)$ is the interaction of two instantons
with radii $\rho_{1,2}$, separation $R$ and relative orientation $U$.
In the sum ansatz, both the $II$ and $IA$ interaction give the same
average repulsion $\kappa^2=\frac{27}{4}\pi^2$ \cite{DP_84}. In the
ratio and streamline ansatz, this repulsion is considerably weaker.
In the streamline ansatz, only the $IA$ interaction is repulsive with
$\kappa^2=4.772$ \cite{Ver_91}.

    If the variational single instanton distribution $d(\rho)\exp
(-\nu\rho^2/\overline{\rho^2})$ is close to the true distribution
for $\alpha=1$ we can calculate the expectation value in
(\ref{int_coup2}) using the variational one. One finds $\langle
S_{int}\rangle \simeq \langle S_{int}\rangle_{\alpha=0} = N\nu/2$
and the resulting estimate for the partition function is
\be
\label{Z_var}
    Z = \frac{1}{((N/2)!)^2} (V\mu_0)^N \exp(-\frac{N\nu}{2}),
\ee
which agrees with the result derived in \cite{DP_84}. Varying
$F=-1/V\cdot\log Z$ with respect to the density one finds the
expected result $N/V=2\mu_0$.

   Numerical results for different interactions were compared in
\cite{Shu_85,Ver_91}. For the sum ansatz, the variational method
gives $N/V=0.18 \Lambda^4$ with $\bar\rho=0.48 \Lambda^{-1}$, whereas
the streamline ansatz gives $N/V=0.54 \Lambda^4$ and $\bar\rho=0.69
\Lambda^{-1}$. The dimensionless diluteness $f=0.5\pi^2\rho^4 (N/V)$
(the fraction of space-time occupied by instantons) of these ensembles
is $f=0.05$ and 0.60 respectively, to be compared with $f=0.06$
in the original instanton liquid model. The variational method was
extended to light quarks in \cite{DP_85,DP_86,IS_89,NVZ_89b}.
We will study this problem in detail in the next section.

    If correlations among instantons are important, the variational
method is not expected to provide an accurate estimate for the
partition function and other observables. Since the main source
of correlations in the instanton liquid are dynamical quarks, this
issue is particularly important for real QCD with two light
and one intermediate mass flavor. Also, as argued in the introduction,
we expect chiral symmetry to be restored due to the formation of
instanton-antiinstanton molecules. This feature is certainly not
captured by the variational model (at least not in its simplest
form), and we will therefore study the full partition function
numerically using the method introduced in this section.

\section{The instanton ensemble at zero temperature}

   In this section we want to present numerical results obtained from
simulations of the instanton liquid at zero temperature. While the
streamline ansatz provides in principle the ''best" classical interaction,
its derivation relies heavily on conformal symmetry and we do not know
how to extend it to finite temperature. In section 6, we will therefore
give a brief discussion of the ratio ansatz ensemble at zero temperature.

    A general problem with calculations in the interacting instanton
model is the treatment of very close instanton-antiinstanton pairs.
On the one hand, these configurations do not contribute significantly
to physical observables such as the quark condensate, hadronic masses
or the topological susceptibility. Very close instanton-antiinstanton
configurations correspond to perturbative fluctuations and should not
be taken into account as a non-perturbative effect. On the other hand,
the partition function of the instanton liquid introduced in the last
section treats even very close pairs as two independent pseudoparticles.
Especially in the streamline ansatz, which provides very little short
range repulsion, this means that close pairs can contribute significantly
to the free energy of the system. Ideally, one would need a consistent
determination of the space of collective variables for very close
pairs and a subtraction procedure for purely perturbative fluctuations.
Unfortunately, such a method is still missing. There is an
interesting suggestion, defining the instanton interaction
(via the optical theorem) by the cross section for isotropic
multi-gluon production \cite{DP_91}. In this case, the existence
of a short range core in the instanton interaction is related to
the question whether the multi-gluon production cross section at
high energy can reach the unitarity bound. If, as it has been
suggested in \cite{DP_94} and other works on the subject, the
cross section only grows until it reaches a square root suppression
and then turns, the instanton-antiinstanton interaction would decrease
until it reaches a minimum at which $S_{int}=-S_0$, and then show a
short distance repulsive core.

  In practice we have decided to deal with the problem of very close
pairs in the streamline ansatz by introducing a purely phenomenological
short range repulsive core
\be
   S_{\rm core} = \beta_1(\bar\rho) \frac{A}{\lambda^4}
   |u|^2, \hspace{1cm}
 \lambda = \frac{R^2+\rho_I^2+\rho_A^2}{2\rho_I\rho_A}
  + \left( \frac{(R^2+\rho_I^2+\rho_A^2)^2}{4\rho_I^2\rho_A^2}
   - 1\right)^{1/2}
\ee
in both the II and IA interactions. Here $\lambda$ is the conformal
parameter that determines the functional form of the streamline
interaction \cite{Ver_91} and the parameter $A$ controls the strength
of the core. This parameter essentially governs the dimensionless
diluteness $f=0.5\pi^2\rho^4(N/V)$ of the ensemble. The second parameter
of the instanton liquid is the scale $\Lambda_{QCD}$ in the partition
function, which fixes the absolute units. Although $\Lambda_{QCD}$ is
known in principle from applications of perturbative QCD, the accuracy
of these determinations is not very high and quantities like the instanton
density  $(N/V)\sim \Lambda^4_{QCD}$ are very sensitive to the precise
numerical value of the scale parameter. As in lattice calculations, one
may therefore fix the scale using the value of some observable, for
example a hadronic mass, as the basic unit. Which quantity to hold
fixed while comparing different theories, e.g. quenched and unquenched
calculations, is pure convention. We have decided to proceed in a
very simple way and fix our units such that $N/V=1\,{\rm fm}^{-4}$
in all cases. This means that in this work 1fm is, by definition, the
average distance between instantons (at $T=0$). The corresponding instanton
density agrees well with the lattice measurements mentioned
above and corresponds to the canonical value of the gluon condensate,
$\langle\frac{\alpha_s}{\pi}G^2\rangle=(350\,{\rm MeV})^4$.

    We have studied the instanton ensemble for various values of the core
parameter $A$. The results are summarized in table 1. By increasing the
core one obviously makes the ensemble more dilute. However, as the
interaction becomes more repulsive, the density in units of the scale
parameter drops and one has to increase the value of $\Lambda_{QCD}$ in
order to keep the physical density fixed. In practice, we have chosen $A=128$,
so that $\Lambda_{QCD}$ remains below 300 MeV, which is roughly the upper
limit of the experimental uncertainty. As a consequence, the ensemble
is not as dilute as suggested by phenomenology ($n=1\,{\rm fm}^{-4}$
and $\rho=0.33$ fm), but comparable to the lattice result $n=(1.4$-$1.6)
\,{\rm fm}^{-4}$ and $\rho=0.35$ fm \cite{CGHN_94}. Two more important
parameters of the ensemble are the average instanton action $S=(8\pi^2)/
g^2$ and the average interaction $S_{int}/N$ per instanton. For $A=128$
we find $S\simeq 6.4$ and $S_{int}/N\simeq1.0$, showing that the system is
still semiclassical and that interactions among instantons are important,
but not dominant.

   Detailed results of our simulations for $A=128$ are shown in
fig.2-5. The partition function at each instanton density was
determined by generating 5000 configurations with 32 instantons
at 10 different coupling constants $\alpha$. The variational ansatz
(\ref{Z_free}) for the partition function was determined from 600
initial sweeps with the full interaction ($\alpha=1$). The integral
(\ref{int_coup2}) was determined by gradually lowering the coupling to
$\alpha=0$ and then raising it back to $\alpha=1$ (hysteresis method).
The difference in the result between the up and down sweeps provides an
estimate of the error in the integral due to incomplete equilibration.
The average between the up and down sweeps usually provides a good
estimate for the correct result, even if equilibration is slow
(as will be the case close to the phase transition).

  Fig.2a shows the free energy versus the instanton density (in units
of $\Lambda^4$) for the pure gauge theory (without fermions). At small
density the free energy is roughly proportional to the density, but at
larger densities repulsive interactions become important, leading to
a well-defined minimum. We also show the average action per instanton
as a function of density. The average action controls the probability
$\exp(-S)$ to find an instanton, but has no minimum in the range of
densities studied. This shows that the minimum in the free energy is a
compromise between maximum entropy and minimum action.

  Fixing the units such that $N/V=1\,{\rm fm}^{-4}$, we have $\Lambda=270$
MeV and the vacuum energy density generated by instantons is $\epsilon=-526\,
{\rm MeV}/ {\rm fm}^3$. This important quantity is related to the gluon
condensate by the trace anomaly
\be
\label{trace}
\epsilon = -\frac{b}{128\pi^2} <g^2G^2>,
\ee
where $b=\frac{11}{3}N_c$ in the pure gauge case. For a sufficiently
dilute system of instantons\footnote{As mentioned above, the average
interaction is only a 15\% correction to the single instanton
action.}, the gluon condensate is simply proportional to the instanton
density, $\epsilon=-b/4(N/V)=-565\,{\rm MeV}/{\rm fm}^3$. The good
agreement of this number with the energy density determined above
shows that our calculation is consistent with the trace anomaly. For
the variational calculation, this is also the case \cite{DP_84}, as
indeed one would expect for any calculation that does not introduce
any dimensionful scale except for $\Lambda$. Also note that not only
the depth of the free energy, but also its curvature (the instanton
compressibility) is fixed from a general theorem. From standard
thermodynamical relations, the compressibility is related to the
mean square fluctuation of the particle number. This quantity is
fixed from a low energy theorem derived by Novikov et
al.~\cite{NSVZ_79}
\be
\label{let}
\int d^4x\, <g^2G^2(x)g^2G^2(0)>&=& \frac{128\pi^2}{b}<g^2G^2> .
\ee
Similar to the trace anomaly (\ref{trace}), this relation follows
from the renormalization properties of QCD. Saturating (\ref{let})
with a dilute system of instantons, one has $\langle (\Delta N)^2
\rangle = \frac{4}{b} N$, where $\Delta N$ is the mean fluctuation
of the number of instantons in a volume $V$. For the compressibility
of the instanton liquid this implies
\be
\label{comp}
   \frac{\partial^2F}{\partial (N/V)^2} &=& \frac{b}{4}
   \left( \frac{N}{V} \right)^{-1}.
\ee
We have determined the compressibility from our data and find
$3.2(N/V)^{-1}$, to be compared with $2.75(N/V)^{-1}$ from eq.(\ref{comp}).
We have also studied density fluctuations in the interacting instanton
liquid \cite{SV_95,SS_95} and found good agreement with the low
energy theorem (see also \cite{DP_84,NVZ_89}).

   Next we study the effects of light quarks on the instanton ensemble.
A problem that is well known in lattice simulations is the fact that
finite volume effects are more severe in the presence of light fermions.
These ''mesoscopic" effects become important if the mass of the Goldstone
bosons $m_\pi \sim \sqrt{m \Lambda}$ is comparable to the inverse box
size $1/L$. In order to avoid these problems one is often forced to
work with light quark masses that are larger than their physical values.
The calculations reported below were performed with light quark masses
$m_u=m_d=0.1\Lambda$, while the strange quark mass is $m_s=0.7\Lambda$.
If one makes the quark mass smaller then chiral symmetry breaking will
eventually disappear due to finite size effects.

      The free energy as a
function of the instanton density in full QCD is shown in fig.3a. We
find that the free energy looks very similar to the pure gauge case, but
the minimum is shifted to a smaller density $N/V=0.174\Lambda^4$. With
our convention $N/V=1\,{\rm fm}^{-4}$ as above, the scale is given by
$\Lambda = 310$ MeV. This gives the instanton-induced vacuum energy
$\epsilon=-280\,{\rm MeV}/{\rm fm}^3$, which is smaller as compared
to the pure gauge case. In the presence of quarks, the trace anomaly
gives\footnote{Despite the factor 1/4 in front of the quark condensate
contribution, this relation is consistent with the general result
$\langle\bar q_fq_f\rangle=\frac{\partial}{\partial m_f}\log Z$.
The remaining mass dependence of the vacuum energy density comes
from the gluon condensate, $\frac{\partial}{\partial m_f}\langle
g^2G^2\rangle = -\frac{96\pi^2}{b}\langle\bar q_fq_f\rangle$, which is
valid for light quarks.}
\be
\label{trace2}
\epsilon = -\frac{b}{128\pi^2} <g^2G^2>
   +\sum_f \frac{1}{4}m_f<\bar q_f q_f>.
\ee
The main difference as compared to the pure gauge case is that the
Gell-Mann-Low coefficient is changed from $b=11$ to $b=9$ (for $N_f=3$).
Estimating the gluon condensate from the instanton density as above, we
get $\epsilon = -488\,{\rm MeV/fm^3}$. This number does not agree very
well with the value extracted from the free energy. One possible reason
is that interactions are more important in the full ensemble so that the
gluon condensate is not proportional to the instanton density. Another
problem might be connected with the value of the quark mass. As mentioned
above, if one enters the regime of mesoscopic QCD, one may encounter
additional scale breaking not described by the trace anomaly. We indeed
find that the scale anomaly can be saturated with the instanton density
for larger quark masses.

    At the minimum of the free energy we can also determine the quark
condensate (see fig.3c). In full QCD we find $\langle\bar qq\rangle=-
(216 {\rm MeV})^3$, which is in good agreement with the phenomenological
value. This value can also be compared to the quark condensate in quenched
QCD (calculated in the pure gauge configurations, see fig.2c) $\langle\bar
qq\rangle=-(251 {\rm MeV})^3$, showing that light quarks suppress the
quark condensate.

    Additional information about the ensemble is provided by the
distribution of the instanton sizes, Dirac eigenvalues and fermionic
overlap matrix elements. In figures 4 and 5 we present these distributions
for the two cases discussed above. The instanton size distribution shows
the perturbative $\rho^{b-5}$ behavior at small sizes, has a maximum
(at $\rho=0.50\Lambda^{-1}$ in quenched and $\rho=0.65\Lambda^{-1}$ in full
QCD) and falls off for large sizes. The average sizes are $\rho=0.43$ fm
in quenched and $\rho=0.42$ fm in full QCD. The diluteness parameter in
the full ensemble is $f=0.14$, to be compared with $f=0.17$ in the
quenched calculation.

   The distribution of eigenvalues of the Dirac operator, $\hat D
\psi_\lambda=\lambda\psi_\lambda$, contains a lot of useful information
about the spectrum of the theory. We will discuss some of these questions
in more detail below. Here we only mention that the Banks-Casher relation
$\langle\bar qq\rangle=-\rho(\lambda=0)/\pi$ connects the density of
eigenvalues of the Dirac operator near zero with the chiral condensate.
{}From figs.4b and 5b one clearly observes that the instanton liquid
leads to a nonvanishing density of eigenvalues near $\lambda=0$.
The results also show that the presence of light quarks suppresses
the number of small eigenvalues. While the spectrum is peaked towards
small eigenvalues in the purge gauge case, it is essentially flat
in full QCD. This is consistent with the prediction from chiral
perturbation theory \cite{SS_93}
\be
\rho(\lambda)= -\pi \langle\bar qq\rangle +
   \frac{\langle\bar qq\rangle^2}{32\pi^2f_\pi^4}
   \frac{N_f^2-4}{N_f} |\lambda|+ \ldots \, ,
\ee
which is valid for $N_f\geq 2$ (speculations about the Dirac spectrum
in quenched QCD can be found in \cite{DT_95}). The second term is
zero\footnote{This result is connected with the fact that for $N_f>2$
there is a Goldstone boson cut appearing in the scalar isovector
($\delta$ meson) correlator, while there is no $\delta
\to \pi\pi$ decay allowed for two flavors.} for $N_f=2$ and leads to a
singular behavior for $N_f>2$. No such dip is seen in our spectrum,
showing that the result is closer to the $N_f=2$ case. More detailed
studies of the spectrum of the Dirac operator in the instanton liquid
for various number of colors and flavors were reported in \cite{Ver_94}.

   In fig.3c and 4c we also show the distribution of the largest fermionic
overlap matrix elements $T_{IA}$ for each instanton. For each instanton
we select the antiinstanton it has the largest overlap with and plot
the resulting distribution of matrix elements. This distribution is a
measure of the strength of correlations among the instantons. For a
completely random system at the same density as the simulated quenched
and fully interacting ensembles, the average overlap would be
$T_{IA}=0.22 \Lambda$ and $T_{IA}=0.21\Lambda$, respectively. Instead,
the measured distribution have an average $0.23\Lambda$
and $0.33\Lambda$, showing that correlations are not very important
in the quenched ensemble, but play some role in the full ensemble.
A more physical measure of the importance of correlations among instantons
is given by the dependence of hadronic correlation functions on the
different ensembles. We will study this question in some detail in
a forthcoming publication.

\section{The phase diagram of the instanton liquid}

    After fixing the the overall scale at zero temperature it is
now straightforward to extend our calculation to finite temperature.
In a euclidean field theory, finite temperature only enters through the
boundary conditions obeyed by the fields. The gauge fields have to be
periodic in the euclidean time direction with the period given by
the inverse temperature, while fermions are subject to antiperiodic
boundary conditions. The corresponding periodic instanton and fermion
zero mode profiles can be constructed using 't Hooft's multi-instanton
solution \cite{HS_78}. These profiles may then be used to study the
instanton interaction and fermionic overlap matrix elements. For the
ratio ansatz, such a study was performed in \cite{SV_91}. A
parameterization of this interaction is given in appendix C and D.

    The most important qualitative feature of the instanton interaction
at finite temperature is the form of the fermionic overlap matrix
elements. The determinant for one instanton-antiinstanton pair
separated by $\tau$ in euclidean time and $r$ in the spatial direction
is roughly proportional to
\be
\label{detD}
  \det(\hat D) \sim \left| \frac{\sin(\pi T\tau)}{\cosh(\pi T r)}
                    \right|^{2N_f} .
\ee
The form of this interaction is a simple consequence of the periodicity
in the time direction and the fact that fermion propagators are screened
in the spatial direction by the lowest Matsubara frequency $\pi T$.
The interaction clearly favors instanton-antiinstanton pairs (molecules)
that are aligned along the time direction with a separation $\tau=1/(2T)$.
Ilgenfritz and Shuryak \cite{IS_94} proposed that this feature of the
interaction leads to a phase transition in the instanton liquid. In
this phase transition, the system goes from a random liquid to an
ordered phase of instanton-antiinstanton molecules. The transition
was studied in a schematic model in \cite{IS_94,SSV_95}. In the present
work we want to test whether a consistent treatment of the full
partition function of the instanton liquid does indeed lead to the
expected phase transition and perform a quantitative study of bulk
properties of the system near the transition.

    Before we consider the QCD case (with two light and one intermediate
mass flavor) in detail, it is instructive to study the phase diagram of
the instanton liquid for a wider range of theories. In order to cover
many different parameters, we have restricted ourselves to an exploratory
study, in which we do not minimize the free energy of the system, but
consider the instanton ensemble at fixed density. Using the same
convention employed in the last section, this density was chosen to be
$1\,{\rm fm}^{-4}$. In our experience, if the system undergoes a phase
transition, it will do so even if the density is not determined
self-consistently, although at a somewhat different temperature.
Nevertheless, before this point is checked in detail, the results
presented in this section should be considered qualitative.

   We first consider the situation for only one quark flavor. In this
case the only chiral symmetry is the anomalous axial $U_A(1)$, so there
are no Goldstone pions, but only a flavor singlet (the analog of the $\eta'$).
Therefore, one would not expect a chiral restoration phase transition. On the
other hand, a (first order) transition may exist even without a symmetry, and
in principle, the general argument concerning the formation of $IA$ molecules
also applies to the case $N_f=1$. Clearly, fermion induced
correlations become stronger as the number of flavors increases, so
whether they are strong enough for $N_f=1$ to induce a phase transition
has to be studied by performing simulations.

   We have studied the $N_f=1$ system at an instanton density $N/V=1\,
{\rm fm}^{-4}$ for temperatures up $T\simeq 300$ MeV. We have observed
no phase transition in this range, the condensate $\langle\bar qq\rangle$
decreases smoothly and is non-zero even at the highest temperature studied.
Of course, even in pure SU(3) Yang Mills theory (without fermions) lattice
simulations find a strong first order deconfinement phase transition at a
temperature $T\simeq 250$ MeV (where the scale is set by the string tension
or quenched $\rho$ meson mass). The interacting instanton model, however,
does not give confinement, so, naturally, there is no deconfinement
transition.

    The case of two light flavors, $N_f=2$, is quite special. According
to standard universality arguments \cite{PW_84,Wil_92}, the chiral phase
transition is expected to be of second order, with the same critical behavior
as the $O(4)$ Heisenberg magnet in $d=3$ dimensions \footnote{Recently, Kocic
and Kogut \cite{KK_95} have challenged these arguments and suggested that the
transition is indeed second order, but with mean field exponents. Lattice
results (see \cite{KL_94} and the review \cite{deT_95}) seem to verify
that the transition is second order, but are not yet sufficiently accurate
to resolve the issue about the critical exponents.}.
We have simulated $N_f=2$ ensembles for various temperatures and
values of the quark masses. For sufficiently small quark masses
$m<0.15\Lambda$ we find large fluctuations in the condensate at
$T\simeq 150 MeV$. To illustrate this fact, we show a typical time
history of the quark condensate in fig.6a. Every configuration
corresponds to one complete sweep through the instanton ensemble,
with one Metropolis hit performed on every collective variable.
It is clear that the transition is either second order or first
order with a very small barrier between the two phases. In order
to distinguish these two possibilities one would have to perform
a finite size scaling analysis, which goes beyond the scope of
this paper. We will see later that for more flavors there is a
barrier between the two phases. In fig.6b we show the Dirac
spectrum for $T=150$ MeV and $m=5$ MeV. The eigenvalue density
extrapolates to zero at $\lambda=0$, but the slope is very steep
and there are no signs of a gap in the spectrum.

   We will treat the three flavor case in detail in the next section,
and now proceed to the case of several flavors of light quarks.
Let us first mention several theoretical arguments suggesting
qualitative changes in the vacuum structure as $N_f$ increases
beyond some critical value $N_f^{crit}$.
\begin{itemize}
\item  A simple phenomenological observation is the fact that for
$N_f>12$ the number of ``pions" $N_{\pi}=N_f^2-1$ exceeds the number
of quarks and antiquarks $N_{q}=4 N_c N_f$. This prevents the usual
matching of the pion gas at low temperature to the high-$T$ quark-gluon
plasma phase with a positive bag constant. It seems plausible that something
should happen before this point is reached, so $N_f^{crit}<12$.

\item As noticed by a number of authors (see, e.g. \cite{BZ_82}), the second
coefficient $b'$ of the beta function changes sign for smaller $N_f$ than
the first one, $b$. In this window one has the possibility of asymptotic
freedom coexisting with an infrared fixed point.

\item  It was recognized in \cite{Shu_87,Shu_88} that for large $N_f$
the instanton liquid is dominated by instanton molecules, and that
their density is UV divergent (at small radii) if $2b-5<0$, corresponding
to $N_f>12$ (this estimate is slightly modified if interactions are taken
into account \cite{SV_90}). This divergence, however, is a short distance
effect and should not affect physical observables.

\item  Lattice simulations of QCD for large $N_f$ were performed by
Iwasaki et al.~\cite{IKS*94}. They found that for $N_c=2$ the critical
number of flavors for which chiral symmetry is restored is $N_f^{crit}=3$,
while for $N_c=3$ it is $N_f^{crit}=7$. A strange ``bulk transition"
was also observed by the Columbia group for $N_f=8$ \cite{Chr_93}.

\item Finally, significant progress has been made in understanding
$N=1$ supersymmetric generalizations of QCD \cite{Sei_94}. One result
that is of particular interest in the present context is the exact
determination of the critical number of flavors $N_f^{crit}=N_c+1$, where
chiral symmetry is broken for $N_f\leq N_f^{crit}$, but not for $N_f>
N_f^{crit}$.
\end{itemize}

   Let us now report on our results concerning the vacuum structure of
the instanton liquid for many flavors. A useful tool to analyze the
structure of chiral symmetry breaking that was suggested by the
Columbia group \cite{Cha_95} is the ``valence quark mass" dependence
of the quark condensate, defined by
\be
\label{qq_val}
<\bar qq(m_v)> &=&-\int d\lambda\,\rho(\lambda,m)
  \frac{2m_v}{\lambda^2+m_v^2}\, .
\ee
Here, $\rho(\lambda,m)$ is the eigenvalue density of the Dirac operator
calculated from configurations generated with dynamical mass $m$. The
quark condensate in the chiral limit is defined by going to the thermodynamic
limit and then taking $m=m_v\to 0$. In a finite system, however, chiral
symmetry is never broken at nonzero quark mass and one has to perform a
detailed scaling analysis. This analysis amounts to performing many
simulations at various values of the dynamical mass, which is
a very time consuming procedure. Instead, one might get a
good indication of the full behavior by studying the valence mass
dependence of the condensate. At very small $m_v$, the condensate
is proportional to $m_v$, which means that one is sensitive to
induced rather than spontaneous chiral symmetry breaking. At very
large mass $m_v$, the condensate is proportional to $m_v^{-1}$.
Both of these limits are of course unphysical, and spontaneous
symmetry breaking in the continuum limit is indicated by the
appearance of a plateau at intermediate values of the valence
mass where $\langle\bar qq\rangle$ depends only weakly on $m_v$ .

   In fig.7 we show a number of calculations of the valence mass dependence
of the condensate. The results in fig.7a were obtained for 64 instantons
in a cubic box $V=(2.828\,{\rm fm})^4$ (corresponding to a density $N/V
=1\,{\rm fm}^{-4}$ and a fairly low temperature, $T=71$ MeV) with the
dynamical quark mass $m=20$ MeV and different number of flavors $N_f=1,
\ldots,8$. Regions where the condensate depends only weakly on $m_v$
are clearly seen for $N_f=1,2,3$. These plateaus are clearly absent for
$N_f=5$ or larger: we therefore conclude that the instanton liquid
model has a chirally symmetric ground state in these cases. A more
detailed study of the configurations shows that all instantons are
bound into molecules. The borderline case appears to be $N_f=4$: if
a condensate is present, it is significantly smaller than in QCD and
due to a relatively small random component of the vacuum.

   We have also performed simulations for two colors,  and
obtained a similar picture, but with the critical number of flavors
smaller by one. For $N_c=2$ no clear signal of chiral symmetry
breaking was observed for $N_f\geq 3$. This suggests the empirical
relation $N^{crit}_f=N_c+C$ with $1<C<2$, an amusing coincidence with
Seiberg's results for $N=1$ supersymmetric QCD \cite{Sei_94}.

    The next question we would like to address is the dependence of these
results on the dynamical quark mass $m$. Increasing the dynamical mass, one
decreases the influence of quark determinant. A large quark mass works
against the correlations induced by a large number of flavors. This means
that for a sufficiently large quark mass one can find ''spontaneous"
symmetry breaking even if there is no symmetry breaking in the chiral
limit. This is demonstrated in Fig.7b, where we show a series of calculations
for $N_f=5$ with the dynamical quark mass $m$ ranging from 20 to 100 MeV.
In this case a plateau in the valence mass dependence of the condensate
reappears at rather small critical quark mass of about 35 MeV.

   Repeating these studies for different numbers of flavors and varying
the mass and temperature, one can  map the phase diagram of the instanton
liquid in the $T-m$ (temperature-mass) plane. In fig.8 we have summarized
our results for $N_f=2,3$ and 5. Those cases in which we found a signal
for chiral symmetry breakdown are marked by the open squares, and those
corresponding to the restored phase by solid squares. The (approximate)
location of the discontinuity line between the two phases is marked by
the stars connected by dashed lines. For two flavors we do not find such
a line of discontinuities. We have performed a number of runs in the
vicinity of the phase transition, all showing large fluctuations of the
condensate. For larger $N_f$ we clearly see a transition line with a
discontinuity of the condensate. When the number of flavors increases,
one end of this line moves to the left (the critical temperature $T_c$
decreases with $N_f$), crossing zero somewhere around $N_f=4$, but before
$N_f=5$. In these cases, the ground state exhibits spontaneous symmetry
breaking only if the quark mass exceed some critical value. For technical
reasons we have not tried to follow the discontinuity lines much beyond
$T>200$ MeV. In this regime, the condensate becomes more and more dominated
by very few extremely small eigenvalues, so that numerical accuracy starts
to become a concern. In addition to that, one does not expect to be
able to describe possible transitions at very large temperature in the
instanton model, since these will presumably be dominated by the pure
gauge deconfinement transition.

\section{The instanton ensemble in QCD at finite temperature}

  In this section we want to discuss in detail the physically
relevant case of two light and one intermediate mass flavor. In
particular, we will perform selfconsistent simulations where the
correct instanton density is determined from minimizing the free
energy. The instanton interaction and fermionic matrix elements
that enter these calculations have already been discussed in the
last section. The semiclassical calculation that leads to the
instanton distribution (\ref{idens}) has also been generalized to
finite temperature \cite{PY_80}, giving
\be
\label{pis}
  d(\rho,T) &=& d(\rho,T=0) \exp(-\frac{1}{3}(2N_c+N_f)
    (\pi\rho T)^2- B(z) ) \\
   B(z) &=& \left(1+\frac{N_c}{6}-\frac{N_f}{6}\right)
    \left(-\log \left( 1+\frac{z^2}{3} \right)
    + \frac{0.15}{(1+0.15 z^{-3/2})^8} \right) \nonumber
\ee
with $z=\pi\rho T$. This correction factor leads to an exponential
suppression of large instantons at high temperature. Its origin
is mainly the scattering of thermal gluons on the instanton. This
phenomenon is related to the Debye mass, which in fact has the same
dependence on $N_c$ and $N_f$. The applicability of (\ref{pis}) is
controlled by two separate conditions, $\rho\ll\Lambda^{-1}$ to ensure
the semiclassical treatment, and $T\gg\Lambda$ to justify a
perturbative treatment of the heat bath. It was therefore argued
in \cite{SV_94} that one should not use the perturbative suppression
factor (\ref{pis}) at temperatures below the phase transition.
This suggestion was indeed verified by lattice simulations for pure
gauge theory \cite{CS_95}, which found a very week temperature dependence
of the instanton density below $T_c$, and an exponential suppression of
instantons consistent with (\ref{pis}) above $T_c$. In practice we have
determined the phase transition temperature without the suppression factor
(\ref{pis}). The final simulations were then performed with an additional
temperature dependent factor $(1-\tanh(\frac{T-T_c} {\Delta T}))/2$
inserted in the exponential appearing in eq.~(\ref{pis}). From our
simulations, we  have determined $T_c\simeq 0.75\Lambda$. We have
taken $\Delta T\simeq 0.2\Lambda$. This number characterizes the
temperature range over which the Debye mass reaches its perturbative
value and is roughly consistent with the result of lattice simulations
\cite{EKS_89}. In our simulations, the value of $\Delta T$ determines
how fast the instanton density drops above $T_c$. On the other hand, we
have checked that the temperature dependence of the quark condensate
is almost independent of this parameter.

   With all the ingredients of the partition function fixed, we
can proceed to our simulations. The free energy is determined as
described in sections 3 and 4. There are no difficulties in principle
associated with applying the coupling constant integration method
at finite temperature. In practice, however, some care is required
since in case the system undergoes a phase transition, the ensemble
at the full coupling $\alpha=1$ is in a different phase as compared
to the random distribution at $\alpha=0$. This implies that the
transition will occur at some value of $\alpha$ during the coupling
constant integration. The hysteresis method described in section 4
is particularly well suited to handle a situation like this.
Physically, this simply means that the variational ansatz (\ref{Z_free}),
while it is still used as a reference distribution, ceases to be
a good approximation to the fully interacting instanton distribution.
In the whole range of temperatures studied, we find that the behavior
of the free energy as a function of the instanton density looks
qualitatively similar to the zero temperature result shown in
figs.2 and 3. In particular, the free energy is negative at small
density and has one fairly well defined minimum. This means that the
phase transition is either smooth or a first order transition with
a small jump in the instanton density, which we cannot resolve due
to finite size effects.

   The resulting instanton density, free energy and quark condensate
are shown in fig.9. In the ratio ansatz the instanton density at
zero temperature is given by $N/V=0.69\Lambda^4$. Taking the density
to be $1\,{\rm fm}^{-4}$ at $T=0$ fixes the scale parameter $\Lambda=222$
MeV and determines the absolute units. The temperature dependence of
the instanton density is shown in fig.9a. It shows a slight increase
at small temperatures\footnote{The zero temperature point corresponds
to a simulation with zero temperature matrix elements in a periodic box
$V=(2.828\Lambda^{-1})^4$, so it is not a true zero temperature calculation.},
starts to drop around 115 MeV and becomes very small for $T>175$ MeV. The
free energy closely follows the behavior of the instanton density. This
means that the instanton-induced pressure first increases slightly, but
then drops and eventually vanishes at high temperature. This behavior
is expected for a system of instantons, but in a complete theory with
perturbative effects, the pressure should always increase as a function
of the temperature. In QCD, the pressure at high temperature is of course
provided by the black body contributions from quarks and gluons. We will
come back to this issue at the end of the section.

   The temperature dependence of the quark condensate is shown in fig.9c.
At temperatures below $T=100$ MeV it is practically temperature independent.
It then starts to drop fast and becomes very small around the critical
temperature $T\simeq 140$ MeV. Note that at this point the instanton
density is $N/V=0.6\,{\rm fm}^{-4}$, slightly more than half the zero
temperature value. This means that the phase transition is indeed caused
by a transition within the instanton liquid, not by the disappearance
of instantons. This is illustrated by fig.10, which shows projections
of the instanton liquid at temperatures $T=74,123,158$ MeV, below, near
and above the chiral phase transition. The figures are projections of
a 4-dimensional cube $V=(3.00\Lambda^{-1})^3\cdot 1/T$ into the 3-4
plane. The positions of instantons and antiinstantons are denoted by
$+/-$ symbols. The lines connecting them indicate the strength of the
fermionic overlap matrix elements. Below the phase transition, there
is no clear pattern. Instantons are unpaired, part of molecules or larger
clusters. As the phase transition progresses, one clearly observes the
formation of polarized instanton-antiinstanton molecules.

   More details of the phase transition can be inferred from the
graphs in figs.11,12 and 13. Fig.11 shows trajectories of the quark
condensate for the temperatures mentioned above. Every configuration
corresponds to one complete sweep through the instanton liquid, with
one Metropolis hit on the collective variables of all instantons. As
discussed in more detail in the next section, the fluctuations in the
quark condensate determine the scalar susceptibility. Below the phase
transition, these fluctuations are controlled by the effective $\sigma$
meson mass. Near the phase transition, we find evidence for a weak first
order transition with the system frequently tunneling between the two
phases. In order to clearly distinguish this scenario from a smooth
crossover we would have to run with significantly larger volumes. Fig.12
shows the spectrum of the Dirac operator. Below the phase transition it
has the familiar flat shape near the origin and extrapolates to a nonzero
density of eigenvalues at $\lambda=0$. Near the phase transition the
eigenvalue density appears to extrapolate to 0 as $\lambda\to 0$, but
there is a spike in the eigenvalue density at $\lambda=0$. This spike
contains the contribution from unpaired instantons. Near the phase
transition, some of these instantons come from the breakup of molecules
as the systems tunnels from the high to the low temperature phase.
In addition to that, even at very high temperature one expects a finite
concentration $O(m^{N_f})$ of random instantons. We will further
comment on this issue in the next section.

   If there is evidence that the transition is weakly first order, it
is of interest to a least have an upper bound on the latent heat
associated with the transition. We have already mentioned that we
do not directly observe a latent heat from a discontinuity in the
instanton density. Another way to calculate the latent heat is
using the Clausius-Clapeyron relation generalized to QCD \cite{Leu_92}
\be
\label{disc}
{\rm disc}\,\epsilon = T_c\left(\frac{\partial T_c}{m_q}\right)^{-1}
 {\rm disc} <\bar qq>\,.
\ee
As Leutwyler already realized in his original paper, this relation
is useful in analyzing numerical simulations even if one does not
directly observe a jump in $\epsilon$. We have determined the
derivative of the critical temperature with respect to the quark
mass from simulations at several different quark masses (see section
7). We find $\partial T_c/\partial m_q\simeq 0.9$ at $m_q=0.1\Lambda$,
close to Leutwyler's estimate $\partial T_c/\partial m_q\simeq 1$. From
the trajectories of the quark condensate (fig.11) one concludes that
the discontinuity is at most ${\rm disc}\langle\bar qq\rangle
\leq 0.4\langle\bar qq\rangle_{T=0}$. These numbers give ${\rm
disc}\,\epsilon \leq 60\,{\rm MeV/{fm}^3}$, which is quite small as
compared to the energy density at zero temperature. Saturating the
trace anomaly $(\epsilon-3p)/4=-\frac{b}{4}(N/V)$ with instantons,
one can also translate this estimate into a bound on the jump in
the instanton density, ${\rm disc}\,(N/V)\leq 0.03\,{\rm fm}^{-4}$,
to be compared with the zero temperature value $(N/V)=1\,{\rm fm}^
{-4}$. This bound shows that we do not expect to see an appreciable
jump in the instanton density, even if there is a first order
transition.

   In fig.13 we show the distribution of $\cos^2(\alpha)=|u_4|^2/|u|^2$,
the relative color orientation angle between instantons and antiinstantons.
For this purpose we have selected for every instanton the antiinstanton
it has the largest overlap with and calculated the corresponding
orientation angle $\cos^2(\alpha)$. The resulting distribution was
normalized to the $SU(3)$ invariant measure. Below the phase transition
there is no preferred direction and the distribution is flat (the
spike near $\cos^2(\alpha)=1$ comes from the fact the measure goes to
zero in this region). Near the phase transition the distribution
is strongly peaked towards $\cos^2(\alpha)=1$, showing that instanton
antiinstanton pairs are indeed strongly polarized.

   Completing this section, let us add a few general remarks concerning
the thermodynamics of the instanton liquid. In this paper, we have only
determined the free energy associated with instantons. Our calculation
includes the contributions from the zero mode determinant, which describes
the excitation of low energy collective modes (the pions). However, it does
not include non-zero mode contributions from quarks or gluons. In particular,
it does not include the Stefan-Boltzmann contribution at large temperature,
or perturbative $O(\alpha_s)$ corrections to it. In a more realistic
description of the thermodynamics of the chiral phase transition, these
contributions should certainly be included. (An attempt to do so was made
in the schematic cocktail model by Ilgenfritz and Shuryak \cite{IS_94}.)
In this paper, however, it was our intention to study the physics
contained in the partition function of the instanton liquid, without
making further assumptions or inputs. Remarkably, the instanton liquid
undergoes a phase transition in the expected temperature regime even
if no additional ingredients are added.

\section{Dirac eigenvalues and susceptibilities}

   Beginning with the classical paper by Banks and Casher \cite{BC_80},
many useful relations have been pointed out between various observables
and the spectrum of the Dirac operator $\hat D \psi_\lambda= \lambda
\psi_\lambda$. In quantum field theories the region of small
$\lambda \rightarrow 0$ is the analog of the Fermi surface in solid
state physics, and understanding the density of states near this point
is crucial for many properties of the system, both at zero and
non-zero temperatures.

    Before we discuss our data in more detail we would like to briefly
review some of the results that can be found in the literature.
As $T\rightarrow T_c$ the quark condensate vanishes. If there is a
second order phase transition for $N_f=2$, the mass dependence of the
condensate is governed by the critical index $\delta$
\be
\left.<\bar q q>\right|_{T=T_{max}} \sim m^{1/\delta},
\ee
where $T_{max}$ denotes the pseudocritical temperature, corresponding
to the position of the peak in the scalar susceptibility (see below).
As mentioned above, there is still some controversy concerning the
values of the critical indices. According to standard universality
arguments, QCD with $N_f=2$ case is analogous to the $O(4)$ Heisenberg
magnet, which has $1/\delta=0.21$. On the other hand, mean field
scaling would give $1/\delta=1/3$.

  Using the Banks-Casher relation
\be
\label{bc}
<\bar q q(T)>&=&-\int d\lambda \rho(\lambda,m,T)
     \frac{2m}{\lambda^2 +m^2}
\ee
one can try to convert this relation into a prediction for the
spectral density of the Dirac operator. The problem is that there
are two sources for the mass dependence of the condensate, the
explicit mass term in the integral (the ''valence mass") and the
implicit mass dependence of the spectral density (the ''sea mass").
Combining these two effects, one can only conclude that
\be
\rho(\lambda,T=T_{max}) \sim \lambda^{1/\delta-\kappa}m^\kappa,
\ee
where $\kappa$ characterizes the explicit mass dependence of the
spectral density. Still, one may hope that the influence of the sea
mass is small and study the behavior of the spectral density near
zero at the phase transition. In fig.6 we have shown the distribution
of eigenvalues at $T_c$ for the case of two flavors. The spectrum
behaves roughly like $\rho(\lambda)\sim\lambda^{0.3}$. In fig.12 we
have displayed the analogous result for QCD. Here, the spectrum
consists of two parts, a smooth part behaving like $\rho(\lambda)
\sim\lambda^2$ and a spike near zero. There are two effects that
contribute to the spike near zero. One is the fact that near a
first order phase transition both the broken and the restored phase
contribute to the spectrum. The other is that at finite mass there
will always be a finite $O(m^{N_f})$ number of unpaired instantons,
giving a contribution of the form $\rho(\lambda)\sim m^{N_f}\delta
(\lambda)$ to the Dirac spectrum.

   Additional information about the phase transition is provided
by mesonic susceptibilities. We define these susceptibilities as
the integral of the corresponding mesonic correlation function
\be
\label{sus}
 \chi_\Gamma = \int d^4x <\bar q(x)\Gamma q(x)
      \bar q(0)\Gamma q(0)>,
\ee
where $\Gamma$ is a spin-isospin matrix with the appropriate
quantum numbers. They characterize the response of the system
to slowly varying external perturbations. For example, the
scalar-isoscalar susceptibility can also be defined as the
second derivative of $\log Z$ with respect to the quark
mass. Near a second order phase transition, the susceptibilities
associated with order parameter fluctuations are expected to
diverge in the chiral limit in a universal manner. In particular,
we have the predictions \cite{Wil_92}
\be
\label{cumul}
   \left.\chi_\sigma\right|_{T=T_{max}} \sim m^{1/\delta-1},
   \hspace{1.5cm}
   \left.\frac{\chi_\sigma}{\chi_\pi}\right|_{T=T_c} = 1/\delta .
\ee
Above the phase transition, chiral symmetry is restored and one has
$\chi_\sigma=\chi_\pi$ as the quark mass is taken to zero. Below
the phase transition, $\chi_\sigma/\chi_\pi\to 0$ as $m\to 0$,
since $\chi_\pi$ has a $1/m$ singularity in the chiral limit,
while $\chi_\sigma$ only has a logarithmic singularity. As noted
by Karsch and Laermann \cite{KL_94}, the second relation in
(\ref{cumul}) is very useful in determining the exponent $\delta$,
since measurements of $\Delta(m,T)=\chi_\sigma/\chi_\pi$ at different
masses are expected to cross at a unique temperature $T=T_c$, where
the value of $1/\delta$ can be extracted from $\Delta(m,T_c)$ (up to
corrections from the non-singular part of the free energy).

   The susceptibilities associated with flavored mesons like the
pion and $\delta$ meson are directly related to the spectrum of the
Dirac operator. Inserting the general decomposition of the quark
propagator in terms of eigenfunctions into the definition
(\ref{sus}), one finds
\be
\label{chipi}
\chi_\pi    &=&  2 \int d\lambda \, \rho(\lambda,m,T)
                    \frac{1}{\lambda^2+m^2} ,\\
\label{chidel}
\chi_\delta &=&  2 \int d\lambda \, \rho(\lambda,m,T)
                    \frac{\lambda^2-m^2}{(\lambda^2+m^2)^2} .
\ee
Comparing (\ref{chipi}) with the Banks-Casher relation (\ref{bc})
one notes that $\chi_\pi=-\langle\bar qq\rangle/m$, a relation that
can also be obtained by saturating the pion correlator with one-pion
intermediate states and using PCAC. The susceptibility in the
scalar-isoscalar ($\sigma$ meson) channel receives an additional
contribution from disconnected diagrams, which can not be expressed
in terms of the spectral density
\be
\label{chisig}
\chi_\sigma = \chi_\delta + 2V\left( <(\bar qq)^2>-<\bar qq>^2 \right).
\ee
The second term measures the fluctuations of the quark condensate in
a finite volume $V$ (as the volume is taken to infinity). Note that
all the susceptibilities (\ref{chipi}-\ref{chisig}) suffer from ultraviolet
divergencies related to the behavior of $\rho(\lambda)$ at large
$\lambda$. Here we are interested in infrared divergencies that
occur when the quark mass is taken to zero, related to the spectral
density at small virtuality $\lambda$. This means that the critical
behavior is unrelated to the ultraviolet behavior which has to be
regularized in some fashion. In our case, the eigenvalue density
is calculated in a basis spanned by the zero modes of the individual
instantons. This means that the high virtuality part of the spectrum
that causes the susceptibilities to diverge is not included anyway.

   One important aspect of the QCD phase transition is the question
whether the $U(1)_A$ symmetry is restored during the transition
\cite{Shu_94}. If this is the case, one expects (for $N_f=2$) two additional
degrees of freedom to become light during the transition, the $\eta'$
and $\delta$. The $\eta'$ susceptibility is connected with fluctuations
of the number of zero modes of the Dirac operator and cannot be measured
directly in a system with total topological charge zero\footnote{It was
demonstrated in \cite{SV_95} that one can study the topological
susceptibility by studying fluctuations of the topological charge
in subvolumes of a large system with total charge zero.}. The $\delta$
susceptibility, however, can easily be studied in our simulations.
If $U(1)_A$ symmetry is restored, we expect $\chi_\delta$ to peak at
the transition and diverge as the quark mass is taken to zero. Furthermore,
above the transition one would have $\chi_\pi=\chi_\delta$ as $m\to 0$.
This also requires that the disconnected part of $\chi_\sigma$ has to
vanish for $T>T_c$ as $m\to 0$.

   We show our results for the mesonic susceptibilities in fig.14 and
15. The results in fig.14 were obtained for three flavors with masses
$m_u=m_d=0.10\Lambda=22$ MeV and $m_s=0.70\Lambda=155$ MeV. They correspond
to the instanton density and free energy discussed in the last section. The
susceptibilities were obtained from 1000 configurations separated by
5 sweeps through an ensemble of 64 instantons. In our simulations, the
physical volume depends on the temperature, since the number of instantons
is held fixed while their density drops with temperature. Near the phase
transition, we have $V_3=(3.5\,{\rm fm})^3$. In fig.15 we show our results
for a slightly smaller light quark mass $m_u=m_d=0.07\Lambda=15$ MeV.

   Fig.14 clearly shows a peak in the sigma susceptibility, indicating
the chiral phase transition. The position of the peak is at $T\simeq
125$ MeV, somewhat too low as compared to the estimate $T_c=140$ MeV
from lattice simulations. The delta susceptibility does not show
any pronounced enhancement. This is also seen from the lower part
of the figure, where we show the disconnected part of the scalar
susceptibility and the difference $\chi_\pi-\chi_\delta$. Clearly,
the peak in $\chi_\sigma$ comes completely from the disconnected
part. Furthermore, $\chi_\pi-\chi_\delta$ becomes very small above
the phase transition, but shows no tendency to go to zero in the
range of temperatures studied. At the smaller mass (see fig.15)
there is a broad enhancement visible in $\chi_\delta$ and $\chi_\pi
-\chi_\delta$ appears to vanish above the phase transition. On
the other hand, if one compares the peak heights of $\chi_\sigma$
for the two different quark masses, one finds $\chi_{\sigma,\,{max}}
\sim m^{-1.9}$, significantly stronger than the ($N_f=2$) universality
prediction. In addition to that, if one calculates the chiral
cumulants $\Delta(m,T)=\chi_\sigma/\chi_\pi$ for the two different
masses, they fail to cross in the regime of temperatures studied.

This may be an indication that the quark mass $m=0.07\Lambda$ is already
too small and one is entering the regime of "mesoscopic" QCD, where the
mass is smaller or comparable to the smallest eigenvalue $\lambda_{min}=
O(1/V)$ of the Dirac operator. There is now a well developed theory of
this regime \cite{LS_92,VZ_93}, which we do not want to enter here.
Instead, we have extended our calculations to larger masses $m_u=m_d=
(0.10,0.13,0.15,0.17)\Lambda$. In fig.16 we show the maximum in the
disconnected part of the scalar susceptibility as a function of the
quark mass. For $m\geq 0.10\Lambda$ the behavior is well described by
a single exponent. Fitting the dependence of the peak height on the
light quark mass, we find $\chi_{dis,\,{max}}\sim m^{-0.84}$, quite
consistent with the ($N_f=2$) universality prediction $\chi_{\sigma,
\,{max}}\sim m^{-0.79}$.

\section{Conclusions}

   We have studied the structure of the instanton liquid at zero
and finite temperature. For this purpose, we have introduced a
method that allows us to extract the free energy of the instanton
liquid from numerical simulations. Using this method, we can
determine the instanton density self-consistently, by minimizing
the free energy.

   We found that for the classical streamline interaction
supplemented by a short range core, the instanton liquid in full
QCD (with two light and one intermediate mass flavor) stabilizes
at a density $N/V=0.174\Lambda^4$. Fixing the scale parameter in
order to reproduce the ''canonical" value $N/V=1\,{\rm fm}^4$
(corresponding to $\langle\frac{\alpha_s}{\pi}G^2\rangle = (360
\,{\rm MeV})^4$), we find a vacuum energy density $\epsilon=-280\,
{\rm MeV}/{\rm fm}^3$ and a quark condensate $\langle\bar qq\rangle
=-(216\,{\rm MeV})^3$. The average instanton size is $\rho=0.42$ fm,
leading to a packing fraction $f=0.5\pi^2\rho^4/(N/V)=0.14$.

   This means that the interacting instanton ensemble, especially
using the ratio ansatz required at finite temperature, is not as
dilute as the random ensemble (where we fixed $N/V=1\,{\rm fm}^{-4}$
and $\rho=0.33$ fm), which so successfully reproduces many hadronic
correlation functions. The situation is quite analogous to the nuclear
matter problem, where reproducing the saturation density and the
binding energy requires a delicate balance between the short range
repulsive core and the intermediate range attraction. In the instanton
liquid, the strength of the core (the short range repulsion) controls
the diluteness of the ensemble. Too much repulsion, however, leads
to a very small total density of instantons, which can only be balanced
by some intermediate range attraction. In the ratio ansatz, the short
range core is weak and the intermediate range attraction fairly shallow,
leading to a rather dense ensemble.

   We note that at this point, the instanton-antiinstanton interaction
at short distances is not a very well defined concept, and the core that
is required to stabilize the ensemble (for the streamline ansatz) is
a purely phenomenological parameter. In practice we have fixed this
parameter in order to reproduce the instanton density and size
distribution inferred from phenomenological considerations and
observed in lattice simulations \cite{CGHN_94,MS_95}. Providing
a more solid theoretical foundation for the instanton interaction
will be an important direction for future study. One possibility,
to derive the short range part of the instanton interaction from
the cross section for isotropic multi-gluon production \cite{DP_91},
was already mentioned in section 4. Another idea, the inclusion of
non-perturbative corrections to the beta function was discussed in
\cite{Shu_95}.

  We have also extended our simulations of the instanton liquid to
finite temperature. For this purpose, we have to replace the instanton
interaction and fermionic overlap matrix elements with their finite
temperature counterparts, satisfying the appropriate boundary
conditions in a euclidean box with temporal extent $1/T$. We have
shown that in full QCD, there is a phase transition at $T\simeq 140$ MeV
in which the chiral symmetry is restored. The mechanism that drives
the transition is the formation of polarized instanton-antiinstanton
molecules. As a result, the number of small eigenvalues associated
with unpaired topological charges is suppressed, and chiral symmetry
is restored. Furthermore, for 2 light flavors the transition appears
to be second order, while for $N_f\geq 3$ there is evidence for a
weak first order transition. We have also investigated the phase
diagram of QCD with many fermion flavors. Qualitatively, it is clear
that many fermion flavors help the formation of molecules. As a result,
the transition temperature drops as the number of light flavors is increased.
For $N_f$ larger than some critical value $N_f^{crit}$, chiral symmetry
is restored even at zero temperature. For $N_c=2,3$ our results are
consistent with $N^{crit}_f=N_c+C$, where $1<C<2$.  We have also studied
the location of the discontinuity of the quark condensate in the $T-m$
plane.

    We have also determined the free energy associated with
instantons, ignoring the perturbative (non-zero mode) contributions
from quarks and gluons. At small $T$ it rises slightly, but then starts
to drop and becomes very small for $T>180$ MeV. While the energy
density and pressure above $T_c$ are dominated by the perturbative
Stefan-Boltzmann contributions, instantons can provide a significant
contribution to the so called ``interaction measure" $B=-(\epsilon-3p)/4$
even at $T>T_c$. Apart from the location of the phase transition, the
most important result of our work is that $B$ retains about half of its
$T=0$ value at $T=T_c$. Using the trace anomaly, the interaction measure
is related to the gluon and quark condensates. Indeed, analyzing lattice
simulations of full QCD one can show that about half of the glue remains
condensed at $T=T_c$ \cite{KB_93,Den_89,AHZ_91}. More specifically,
lattice simulations show that the pressure remains small until $T\simeq 2T_c$,
while the energy density reaches the Stefan-Boltzmann value very quickly
\cite{EFK*90,KSW_91,deT_95}. It was emphasized in \cite{KB_93} that this
behavior requires that only part of the bag pressure (or the gluon
condensate) is removed across the phase transition. This is precisely
the behavior observed in the instanton liquid model.

   Finally, we have studied the behavior of the Dirac spectrum and the
mesonic susceptibilities near the phase transition. This is related to
the question whether the behavior of these quantities is governed by
critical scaling due to a nearby second order phase transition, and,
more generally, what symmetries are restored in the transition. We find
a peak in the scalar-isoscalar ($\sigma$) susceptibility, indicating the
chiral phase transition. The dependence of the peak height on the
light quark mass is consistent with the expected scaling behavior
at fairly large quark masses $m\geq 0.1 \Lambda$, but is more singular
at smaller masses, probably due to the onset of finite size effects.
For $m\geq 0.1\Lambda$, there is no hint of $U(1)_A$ restoration: there
is no peak in the isovector scalar ($\delta$) susceptibility and the
Dirac spectrum contains (approximate) zero modes above $T_c$. At
smaller masses the scalar-isovector ($\delta$) susceptibility shows
an ehancement around the critical point, but with the available data
it is difficult to say whether this behavior will persist for larger
volumes.

\section{Acknowledgements}
  We would like to thank J.~J.~M.~Verbaarschot for many useful
discussion. We have also benefitted from conversations with
A.~Smilga and I.~Zahed. Part of this work was done while T.~S.~was
a visitor at the Institute for Nuclear Theory in Seattle. The
reported work was partially supported by the US DOE grant
DE-FG-88ER40388. Most of the numerical calculations were
performed at the NERSC at Lawrence Livermore.

\newpage
\appendix

\section{Instanton interaction at zero temperature}

    In this appendix we specify the classical instanton interaction
used in our simulations. From the form of the action for the
two-instanton ansatz, one can easily show that the general form
of the instanton-antiinstanton interaction in the case of the
gauge group $SU(2)$ is given by
\be
\label{su2_int}
 S_{int} = \beta_1 (\bar\rho) \left(
  s_0(R)+s_1(R)(u\cdot\hat R)^2+s_2(R)(u\cdot\hat R)^4  \right),
\ee
where $u_\mu$ is the orientation vector introduced in section 2 and
$\hat R_\mu$ is the unit vector connecting the centers of the
instanton and antiinstanton. The interaction is given in units
of the single instanton action $S_0=\beta_1(\bar\rho)$. The argument
of the beta function is not uniquely determined without a higher
order calculation. In practice we have taken the geometric average
$\bar\rho=\sqrt{\rho_I\rho_A}$ of the instanton radii. For large
separation $R$, the interaction (\ref{su2_int}) can be embedded
into $SU(3)$ by making the replacement $(u\cdot\hat R)^2 \rightarrow
|u\cdot\hat R|^2$ and multiplying the first term by $|u|^2$.

   For the ratio ansatz, the instanton-antiinstanton interaction can
be parameterized by \cite{SV_91}
\be
\label{S_IA,ratio}
 \frac{S_{IA}}{\beta_1} &=&
   \left( \frac{4.0}{(r^2+2.0)^2}
   - \frac{1.66}{(1+1.68 r^2)^3}
   + \frac{0.72\log (r^2)}{(1+0.42 r^2)^4}
    \right) |u|^2  \\
   & & \hspace{1cm} +  \left( -\frac{16.0}{(r^2+2.0)^2}
         + \frac{2.73}{(1+0.33 r^2)^3}\right)
         |u\cdot\hat R|^2 , \nonumber
\ee
where $r=R/\sqrt{\rho_I\rho_A}$ is the instanton-antiinstanton separation
in units of the geometric mean of their radii. In the ratio ansatz, the
instanton-instanton interaction is non-zero if the two instantons have
different color orientations. The interaction can be parameterized by
\be
\label{S_II,ratio}
 \frac{S_{II}}{\beta_1} &=&
         \frac{1}{(1+0.43 r^2)^3}
         \left[ 0.63 |\vec u|^2 + 0.071 |\vec u|^4 \right] \\
   & & \hspace{0.6cm} \mbox{} -
         \frac{\log (r^2)}{(1+1.17 r^2)^4}
         \left[ 0.05 |\vec u|^2 + 0.47 |\vec u|^4 \right]
        \nonumber .
\ee
In the streamline ansatz, conformal symmetry dictates that the interaction
depends on the relative separation and the instanton radii only through
the conformal parameter
\be
\label{conf_param}
 \lambda = \frac{R^2+\rho_I^2+\rho_A^2}{2\rho_I\rho_A}
  + \left( \frac{(R^2+\rho_I^2+\rho_A^2)^2}{4\rho_I^2\rho_A^2}
   - 1\right)^{1/2}.
\ee
The instanton-antiinstanton interaction is then given by \cite{Ver_91}
\be
\label{S_IA,Yung}
 \frac{S_{IA}}{\beta_1} &=&
   \frac{4}{(\lambda^2-1)^3}  \bigg\{
    -4\left( 1-\lambda^4 + 4\lambda^2\log(\lambda) \right)
         \left[ |u|^2-4|u\cdot\hat R|^2 \right] \\
   & & \hspace{1cm}\mbox{} +
     2\left( 1-\lambda^2 + (1+\lambda^2)\log(\lambda) \right)
         \left[ (|u|^2-4|u\cdot\hat R|^2)^2
                + |u|^4 + 2(u)^2(u^*)^2 \right]
       \bigg\}\nonumber .
\ee
As discussed in section 2, the instanton-instanton interaction vanishes
in the streamline ansatz.

\section{Fermionic overlap matrix elements at zero temperature}

   A parameterization of the fermionic overlap matrix element
between instantons and antiinstantons in the sum ansatz was
already given in section 2
\be
\label{sum_overl2}
 T_{IA} = i (u\cdot R)  \frac{1}{\rho_I\rho_A}
    \frac{4.0}{(2.0+R^2/\rho_I\rho_A)^2} .
\ee
The matrix element in the ratio ansatz is very similar and we
employ the same parameterization. In the streamline ansatz,
the matrix element depends on the separation and the radii again only
through the conformal parameter $\lambda$. The matrix element can
be parameterized by \cite{SV_92}
\be
\label{stream_over}
 T_{IA} = i (u\cdot R)  \frac{1}{\rho_I\rho_A}
    \frac{c_1 \lambda^{3/2}}
         {(1+1.25(\lambda^2-1)+c_2(\lambda^2-1)^2 )^{3/4}},
\ee
with $c_1=\frac{3\pi}{8}$ and $c_2=(\frac{3\pi}{32})^{4/3}$.

\section{Instanton interaction at finite temperature}

    In this appendix we give a parameterization of the classical
instanton interaction at finite temperature. For reasons explained
in section 4, there is no analog of the streamline interaction
at finite temperature and we only specify the interaction in the
ratio ansatz. Up to small changes that have been introduced in order
to improve the parameterization at small temperatures, the interaction
is identical to the one given in \cite{SV_91}. Here we also specify how
to embed the $SU(2)$ parameterization into $SU(3)$.

    At finite temperature, the interaction is still at most quartic
in the relative orientation vector $u_\mu$. However, since four
dimensional rotational invariance is broken, the interaction depends
on $|u_4|^2$ in addition to the invariants $|u\cdot\hat R|^2$ and
$|u|^2$ appearing in the zero temperature interaction. For the
same reason, the interaction can now depend separately on the
spatial and temporal components of the vector $R_\mu=(\vec R,R_4)$.
At temperatures of interest here, the anisotropy in the dependence
on $R_\mu$ turns out to be small. The parameterization
\be
\label{S_IA,finiteT}
 \frac{S_{IA}}{\beta_1} &=&
   \frac{4.0}{(r^2+2.0)^2} \frac{\beta^2}{\beta^2+5.21}|u|^2
   - \left( \frac{1.66}{(1+1.68 r^2)^3}
   + \frac{0.72\log (r^2)}{(1+0.42 r^2)^4} \right)
     \frac{\beta^2}{\beta^2+0.75} |u|^2  \\
   & & \hspace{0.6cm}\mbox{} +  \left( -\frac{16.0}{(r^2+2.0)^2}
         + \frac{2.73}{(1+0.33 r^2)^3}\right)
         \frac{\beta^2}{\beta^2+0.24+11.50 r^2/(1+1.14 r^2)}
         |u\cdot\hat R|^2
         \nonumber \\
   & & \hspace{0.6cm}\mbox{} +
         0.36\log \left( 1+\frac{\beta}{r}\right)
         \frac{1}{(1+0.013 r^2)^4}
         \frac{1}{\beta^2+1.73}
         (|u|^2-|u\cdot\hat R|^2-|u_4|^2)
         \nonumber
\ee
therefore depends only on $r=R/\sqrt{\rho_I\rho_A}$ with $R=(\vec R^2
+R_4^2)^{1/2}$. The inverse temperature $\beta=1/(T\sqrt{\rho_I\rho_A})$
is also given
in units of the mean of the instanton radii. One can easily check that
for $T\rightarrow 0$ the interaction (\ref{S_IA,finiteT}) reduces to
the zero temperature ratio ansatz (\ref{S_IA,ratio}). The interaction
between two instantons can be parameterized by
\be
\label{S_II,finiteT}
 \frac{S_{II}}{\beta_1} &=&
         \frac{1}{(1+0.43 r^2)^3}
         \frac{\beta^2}{\beta^2+5.33}
         \left[ 0.63 |\vec u|^2 + 0.071 |\vec u|^4 \right] \\
   & & \hspace{0.6cm}\mbox{} -
         \frac{\log (r^2)}{(1+1.17 r^2)^4}
         \frac{\beta^2}{\beta^2+1.17}
         \left[ 0.05 |\vec u|^2 + 0.47 |\vec u|^4 \right]
        \nonumber  \\
   & & \hspace{0.6cm}\mbox{} +
       \log\left(1+\frac{\beta}{r}\right)
       \frac{1}{\beta^2+2.08}
       \left[ 0.07 |\vec u|^2 + 0.05 |\vec u|^4 \right]
        \nonumber
\ee

\section{Fermionic overlap matrix elements at finite
temperature}

   At finite temperature, the fermionic overlap matrix element is
still linear in the relative orientation vector $u_\mu$. Due to the
loss of Lorentz invariance at finite temperature, the matrix element
depends separately on $u_4$ and $\vec u\cdot\vec R$
\be
\label{overl_finiteT}
 T_{IA} = i u_4 f_1 + i\frac{(\vec u\cdot \vec R)}{R} f_2 .
\ee
A parameterization of the functions $f_{1,2}$ was given in \cite{SV_91}.
We have changed this parameterization slightly in order to improve
the behavior at small temperatures. The result for $f_1$ is
\be
\label{f1}
 f_1 &=& \frac{\left(\frac{\nsz\pi}{\nsz\beta}\right)
             \sin\left(\frac{\nsz\pi\tau}{\nsz\beta}\right)
              \cosh\left(\frac{\nsz\pi r}{\nsz\beta}\right)}
              {\left(\cosh\left(\frac{\nsz\pi r}{\nsz\beta}\right)-
               \cos\left(\frac{\nsz\pi\tau}{\nsz\beta}\right)+
               \kappa_1^2\right)^2} \\
     & & \cdot \frac{1}{\left(\frac{\nsz\beta^2}{\nsz\pi^2}\right)
                  \left(\exp\left(-\frac{\nsz\pi r}{\nsz 2\beta}\right)+
                  \frac{\nsz\pi r}{\nsz 2\beta}\right)
              +   \left(\frac{\nsz 2}{\nsz\pi}\right)\left(1-0.69
                  \exp\left(-\frac{\nsz 1.75r}{\nsz\beta} \right)\right)}
              \nonumber \\
     & & \cdot \left( 1 + \frac{0.76\beta^2}
                         {(0.82r^2+1)^2(1+0.19\beta^2)^2} \right)
         \left( 1 + \left(\frac{\pi\tau}{\beta}\right)^2
                 \frac{0.18}{1+0.12r^2} \right) \nonumber,
\ee
where $r=|\vec R|/\sqrt{\rho_I\rho_A}$, $\tau=R_4/\sqrt{\rho_I
\rho_A}$ and $\beta=1/(T\sqrt{\rho_I\rho_A})$ are given in units of
the mean instanton radius. The  result for $f_2$ is
\be
\label{f2}
 f_2 &=& \frac{\left(\frac{\nsz\pi}{\nsz\beta}\right)
             \cos\left(\frac{\nsz\pi\tau}{\nsz\beta}\right)
              \sinh\left(\frac{\nsz\pi r}{\nsz\beta}\right)}
              {\left(\cosh\left(\frac{\nsz\pi r}{\nsz\beta}\right)-
               \cos\left(\frac{\nsz\pi\tau}{\nsz\beta}\right)+
               \kappa_2^2\right)^2} \\
     & & \cdot \frac{1}{\left(\frac{\nsz\beta^2}{\nsz\pi^2}\right)
                  \left(\exp\left(-\frac{\nsz 2.06\pi r}{\nsz\beta}\right)+
                  \frac{\nsz\pi r}{\nsz 2\beta}\right)
              +   \left(\frac{\nsz 2}{\nsz\pi}\right)\left(1+0.42
                  \exp\left(-\frac{\nsz 0.34 r}{\nsz\beta} \right)\right)}
              \nonumber ,
\ee
with $\kappa_{1,2}$ given by
\be
\label{kappa12}
   \kappa_1^2 = \frac{1}
   {0.53+\left(\frac{\nsz\beta^2}{\nsz\pi^2}\right)},
   \hspace{0.5cm}
   \kappa_2^2 = \frac{1}
   {0.69+\left(\frac{\nsz\beta^2}{\nsz\pi^2}\right)}.
\ee
One can easily verify that the parameterization
(\ref{overl_finiteT}-\ref{kappa12}) reduces to
(\ref{sum_overl2}) in the zero temperature limit.


\newpage\noindent

\begin{figure}
\caption{Classical instanton-antiinstanton interaction
in the streamline (dash-dotted line) and ratio ansatz (short dahed
line). The interaction is given in units of the single instantanton
action $S_0$ for the most attractive ($\cos\theta=1$) and most repulsive
($\cos\theta=0$) orientations. The dash-dotted curves show the original
streamline interaction, while the solid curves show the interaction
including the core introduced in section 3. Fig.1b shows the fermionic
overlap matrix elements in the streamline (solid curve) and ratio ansatz
(dashed curve). The matrix elements are given in units of geometric mean
of the instanton radii.}
\end{figure}
\begin{figure}
\caption{   Free energy, average instanton size and quark
condensate as a function of the instanton density in the pure gauge
theory. All quantities are given in units of the scale parameter
$\Lambda_{QCD}$.}
\end{figure}
\begin{figure}
\caption{   Free energy, average instanton size and quark
condensate as a function of the instanton density in the theory with
two light and one intermediate mass flavor. All quantities are given
in units of the scale parameter $\Lambda_{QCD}$.}
\end{figure}
\begin{figure}
\caption{   Distribution of instanton sizes, eigenvalues
spectrum of the Dirac operator and distribution of fermionic
overlap matrix elements in pure gauge theory. $\rho,\lambda$
and $T_{IA}$ are given in units of the scale parameter. The
distribution functions have arbitrary units.}
\end{figure}
\begin{figure}
\caption{   Distribution of instanton sizes, eigenvalue
spectrum of the Dirac operator and distribution of fermionic
overlap matrix elements in full QCD.}
\end{figure}
\begin{figure}
\caption{   Trajectory of the quark condensate (in units
of ${\rm fm}^{-3}$ for $N_f$=2 in the critical region, $T=150$ MeV.
We also show the corresponding spectrum of Dirac eigenvalues $\lambda$
$[{\rm fm}^{-1}]$. The normalization of the eigenvalue distribution
is arbitrary.}
\end{figure}
\begin{figure}
\caption{   Double logarithmic plot of the quark condensate
(in units ${\rm fm}^{-3}$) versus the valence quark mass $m_v$ for
different number of light flavors (with dynamical mass $m=20$ MeV).
Fig.7b shows the same plot for $N_f=5$ and different quark masses $m$.}
\end{figure}
\begin{figure}
\caption{   Schematic phase diagram of the instanton liquid for
different numbers of quark flavors, $N_f$=2,3 and 5. We show the phase
of chiral symmetry breaking in the temperature-quark mass planes. For
$N_f$=2, open squares indicate points where we found large fluctuations
of the chiral condensate, the cross indicates the approximate location
of the singularity. In the two other figures the open squares correspond
to points where we find a plateau in the valence mass dependence of the
chiral condensate, while solid squares correspond to points where such
a plateau is absent. The crosses and the dashed lines connecting them
show the approximate location of the discontinuity line.}
\end{figure}
\begin{figure}
\caption{   Instanton density, free energy and quark condensate
as a function of the temperature in full QCD  with two light and
one intermediate mass flavor.}
\end{figure}
\begin{figure}
\caption{   Typical instanton ensembles for $T=75$, 123 and
158 MeV. The plots show projections of a four dimensional $(3\Lambda^{-1})^3
\times T^{-1}$ box into the 3-4 ($z$ axis-imaginary temperature) plane.
Instantons and antiinstanton positions are indicated by $+$ and $-$ symbols.
Dashed, solid and thick solid lines correspond to fermionic overlap matrix
elements $T_{IA}>0.40,\,0.56,\,0.64$, respectively.}
\end{figure}
\begin{figure}
\caption{   Trajectories of the quark condensate at three
different temperatures $T=75,\,130,\,158$ MeV. The quark condensate is
given in units of the scale parameter.}
\end{figure}
\begin{figure}
\caption{   Spectrum of the Dirac operator for different
temperatures $T=75,\,130,\,158$ MeV. The normalization of the spectra is
arbitrary (but identical).}
\end{figure}
\begin{figure}
\caption{   Distribution of the color orientation angle
$\cos^2\alpha=|u_4|^2/|u|^2$ for different temperatures $T=75,\,130,
\,158$ MeV. The distributions are normalized to the corresponding $SU(3)$
measure. The overall scale is arbitrary.}
\end{figure}
\begin{figure}
\caption{   Mesonic susceptibilities as a function of temperature
for quark masses $m_u=m_d=0.10\Lambda$ and $m_s=0.70\Lambda$. The scalar
$\sigma$, scalar $\delta$ and pseudoscalar $\pi$ susceptibilities are
denoted by solid circles, solid squares and open squares, respectively.
The lower panel shows the disconnected part of the scalar susceptibility
and the difference $\chi_\pi-\chi_\delta$. The solid lines show a smooth
fit to the data.}
\end{figure}
\begin{figure}
\caption{   Mesonic susceptibilities for a smaller light quark
mass $m_u=m_d=0.07\Lambda$. Curves labeled as in figure 14.}
\end{figure}
\begin{figure}
\caption{   Maximum of the disconnected part of the scalar
susceptibility as a function of the quark mass. Also shown is the
parameterization $\chi \sim m^{-0.84}$.}
\end{figure}

\bigskip
\begin{table}
\caption{Parameters of the pure gauge instanton ensemble in the
streamline ansatz for various values of the core parameter $A$.
The density and average instanton size are given in units of
$\Lambda_{QCD}$ and fm. The dimensionless scale parameter $f$
is introduced in the text.}

\begin{tabular}{rcccccc}
 $A$  &   $N/V\,[\Lambda_{QCD}^4]$     &   $\rho\,[\Lambda_{QCD}^{-1}]$  &
 $f$  &   $\Lambda_{QCD}\,[{\rm MeV}]$ &
          $N/V\,[{\rm fm}^{-4}]$       &   $\rho\,[{\rm fm}]$ \\ \tableline
  24  &        0.97                    &       0.60                      &
0.62  &         202                    &
               1.00                    &       0.59            \\
  64  &        0.41                    &       0.58                      &
0.16  &         250                    &
               1.00                    &       0.45            \\
 128  &        0.30                    &       0.58                      &
0.12  &         270                    &
               1.00                    &       0.43            \\
 256  &        0.18                    &       0.55                      &
0.08  &         307                    &
               1.00                    &       0.36
\end{tabular}

\end{table}

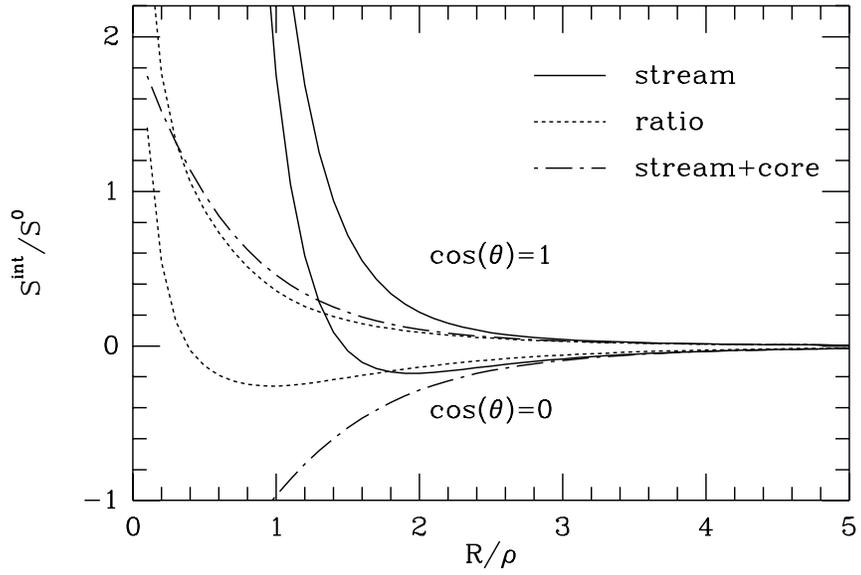
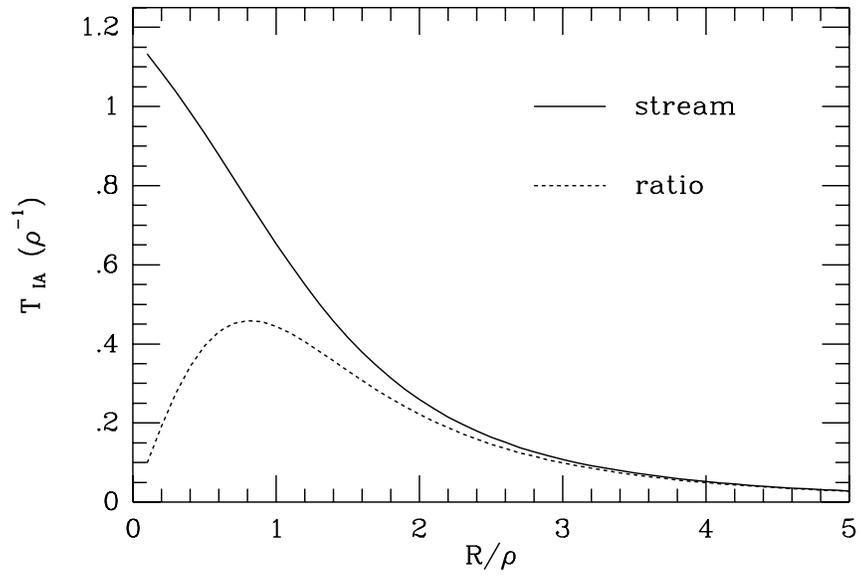
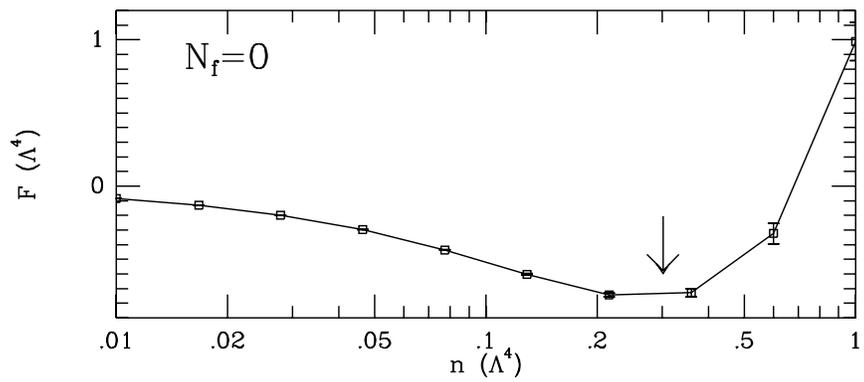
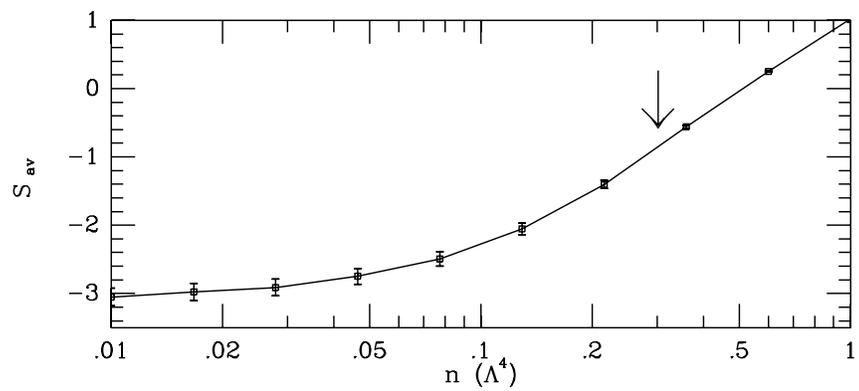
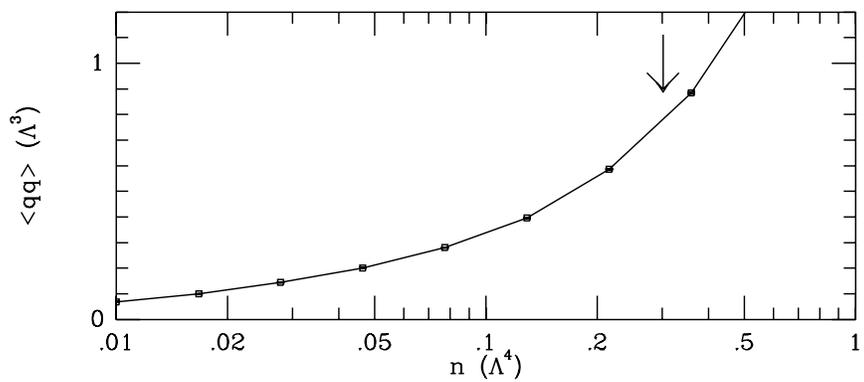
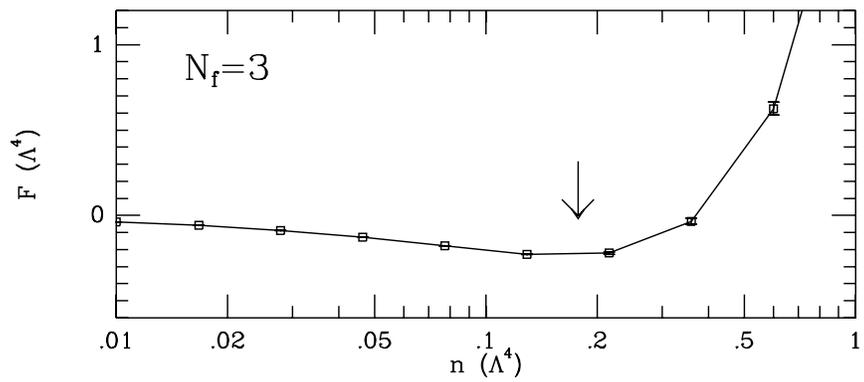
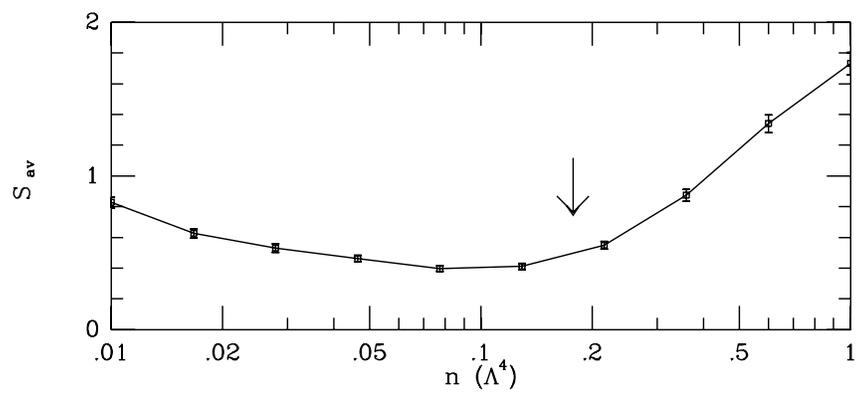
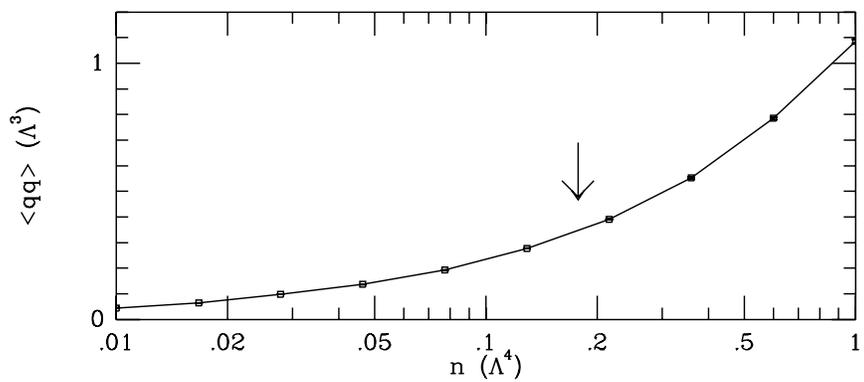
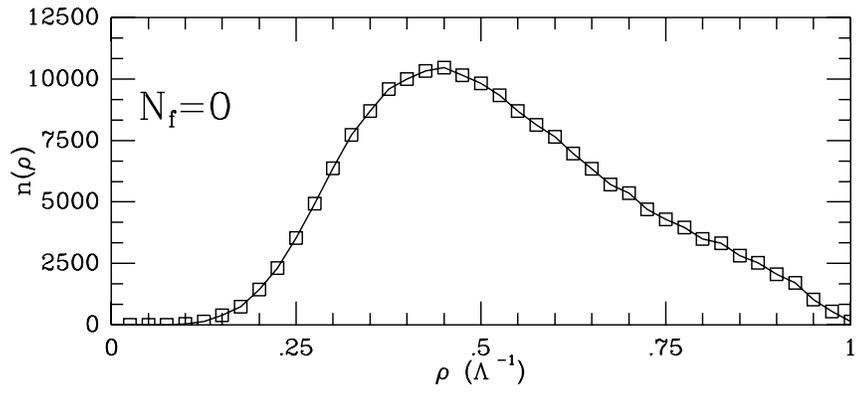
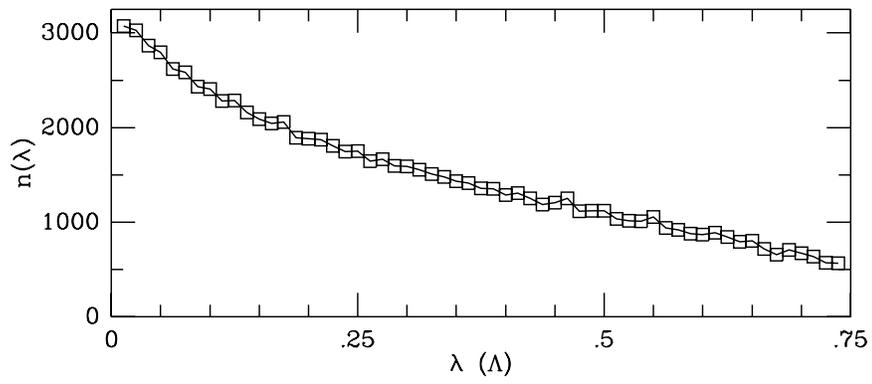
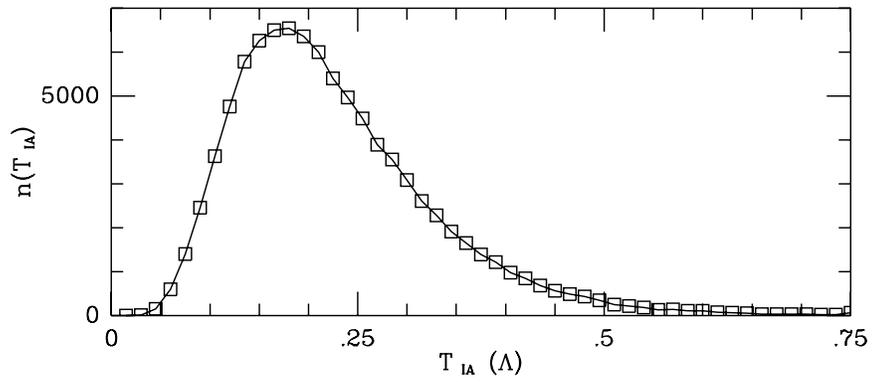
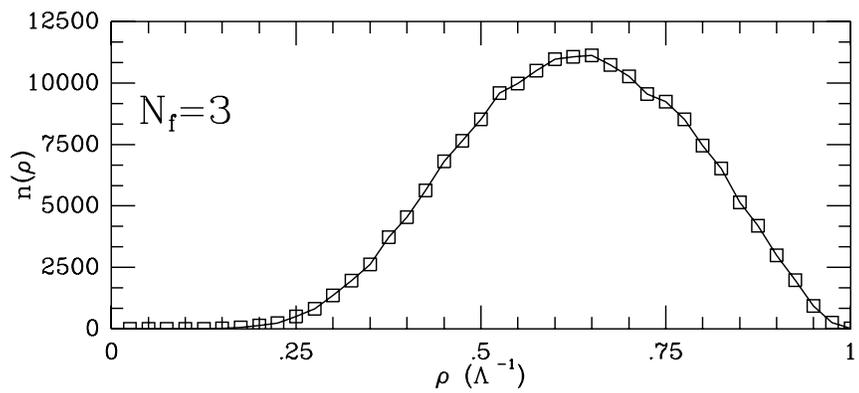
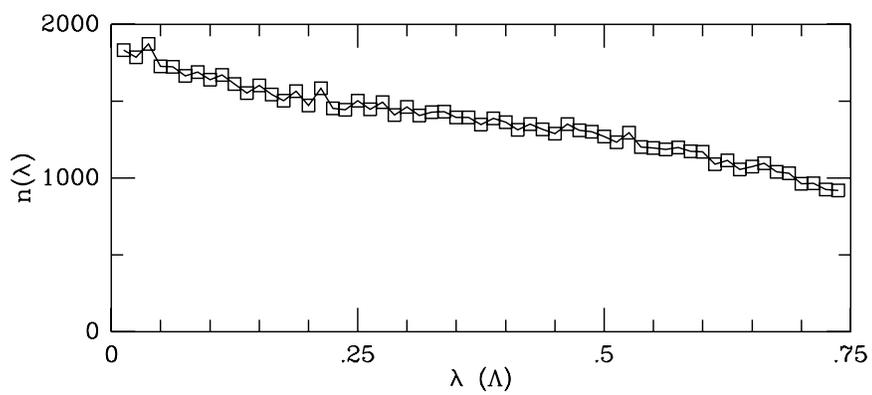
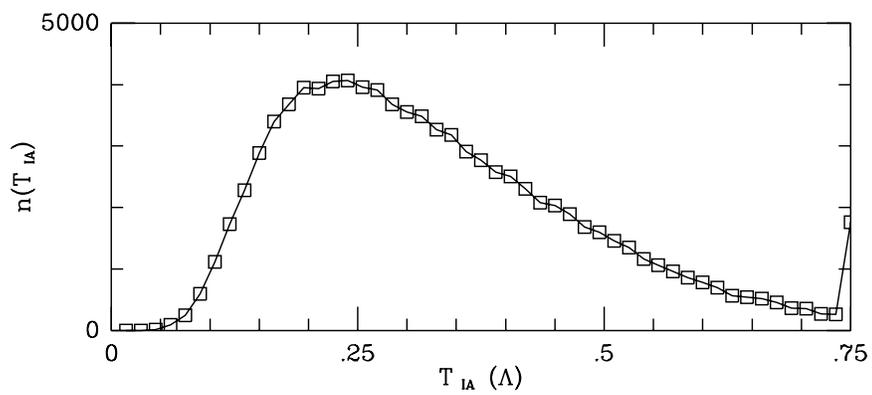
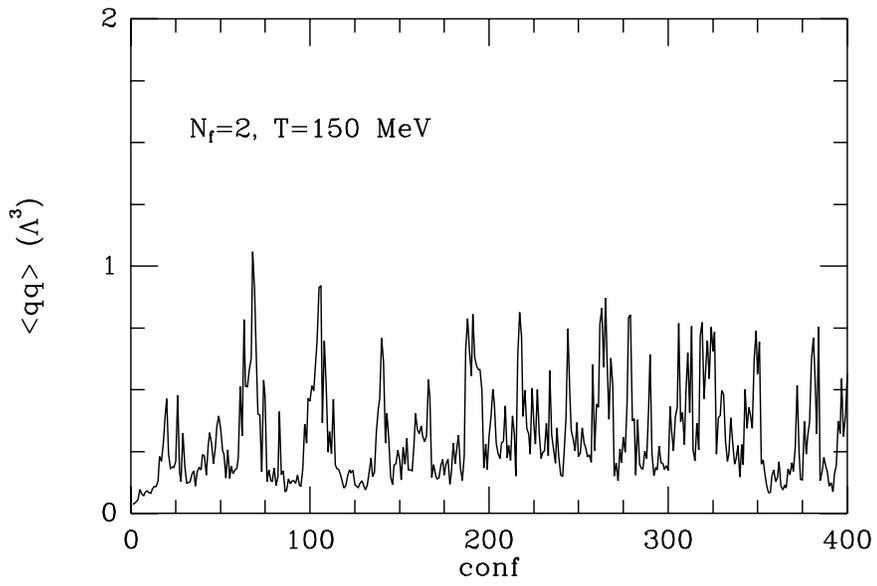
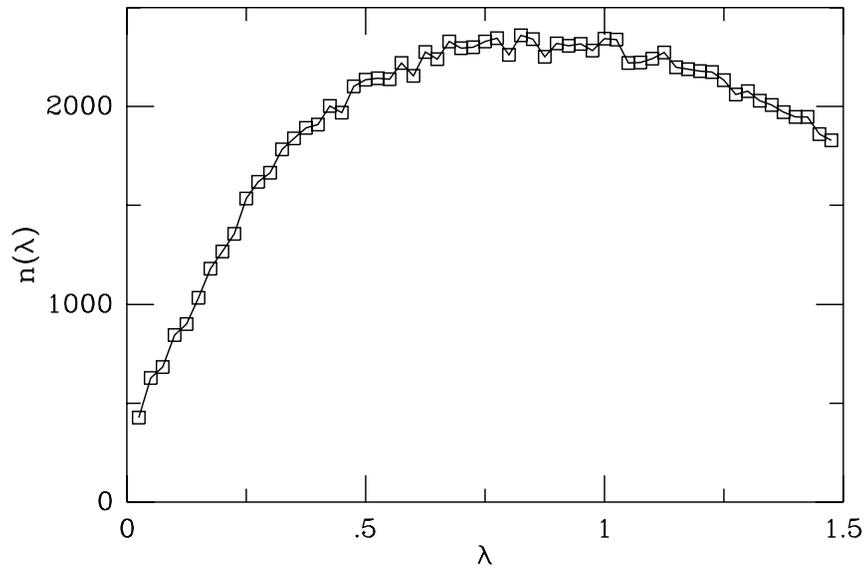
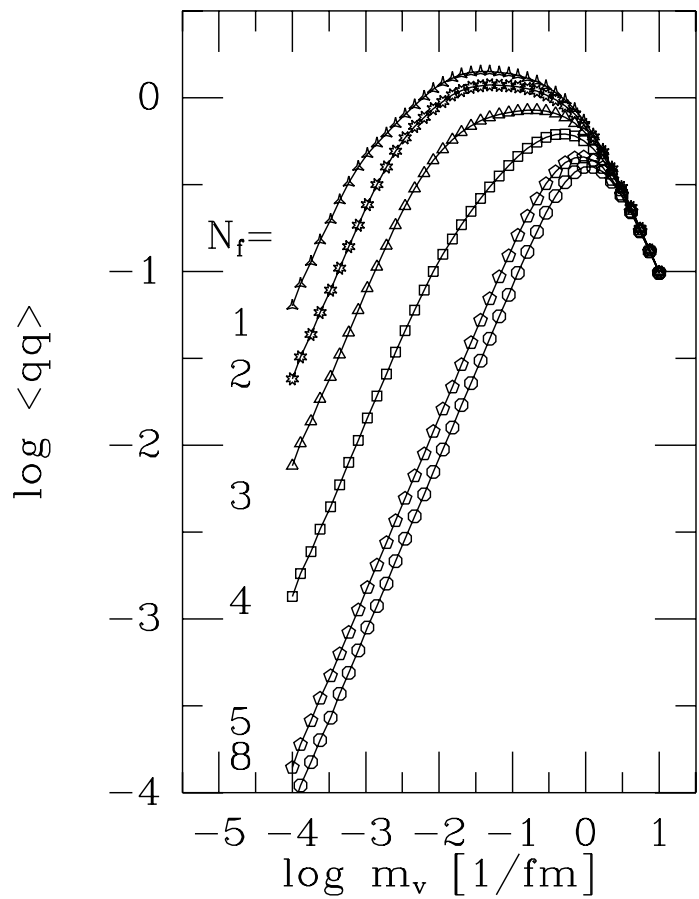
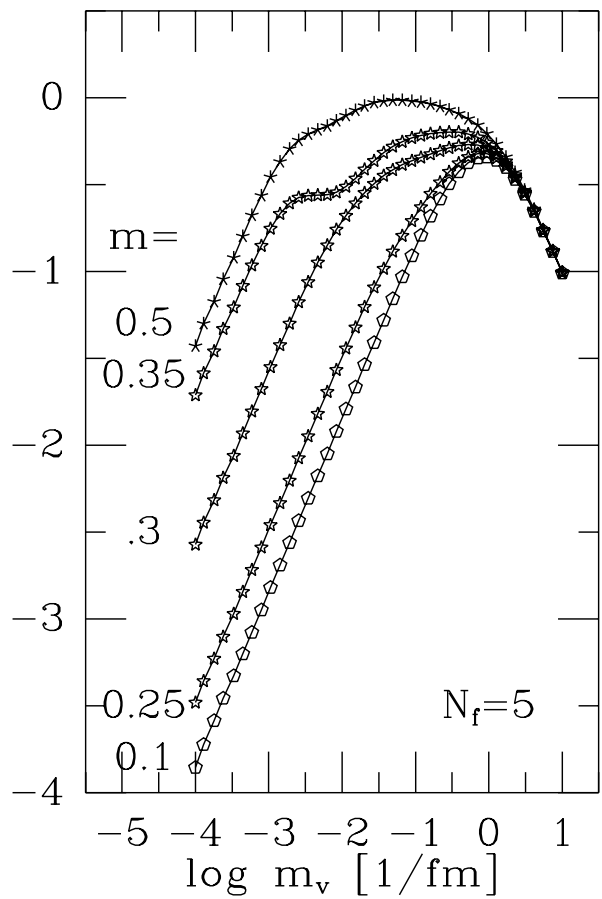
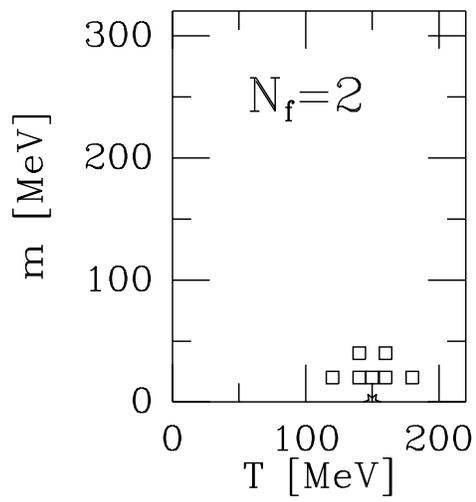
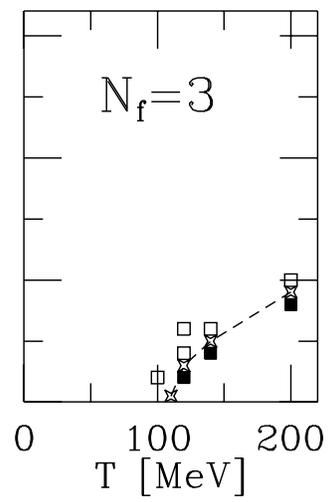
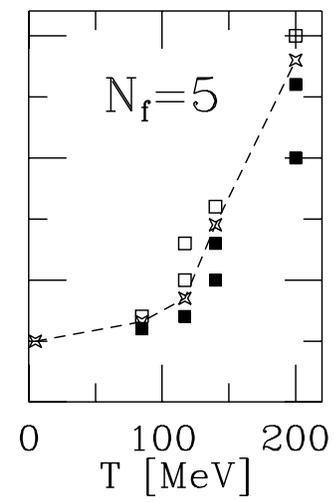
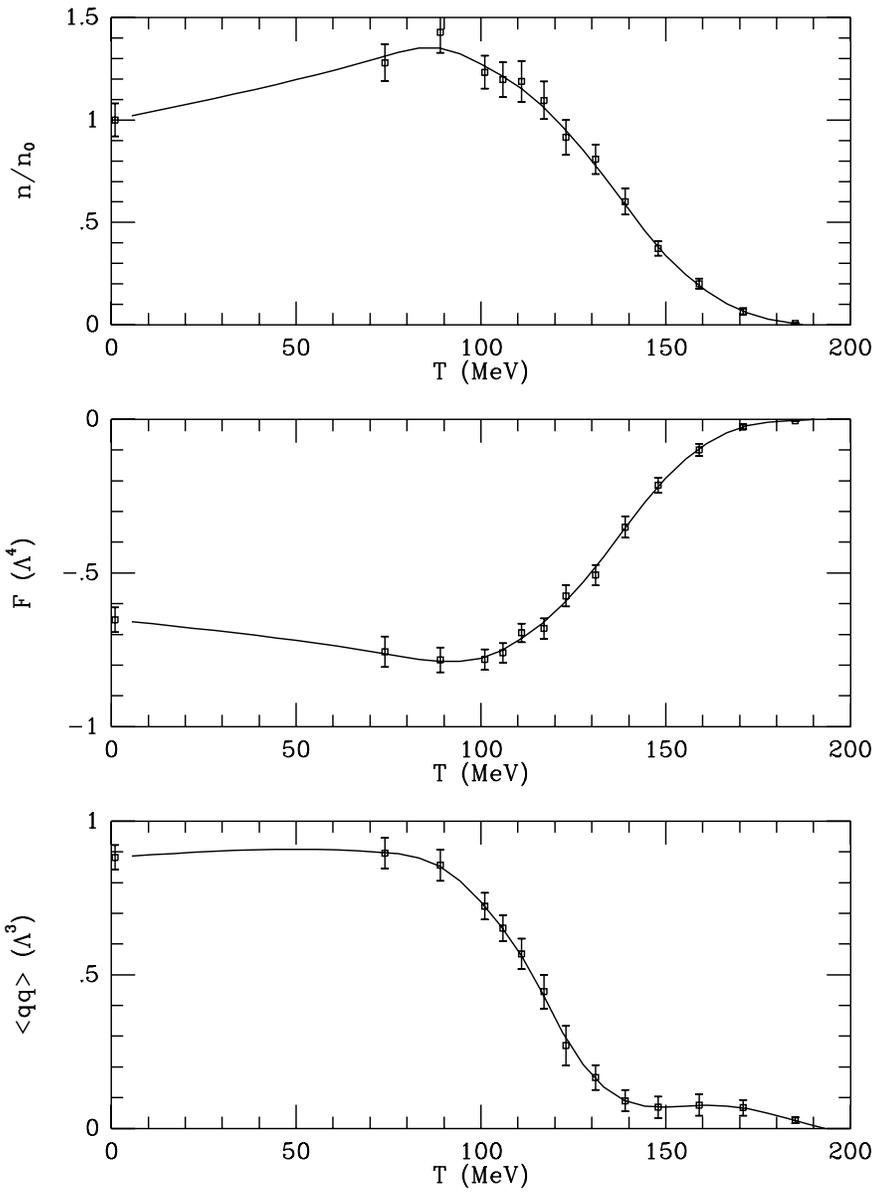
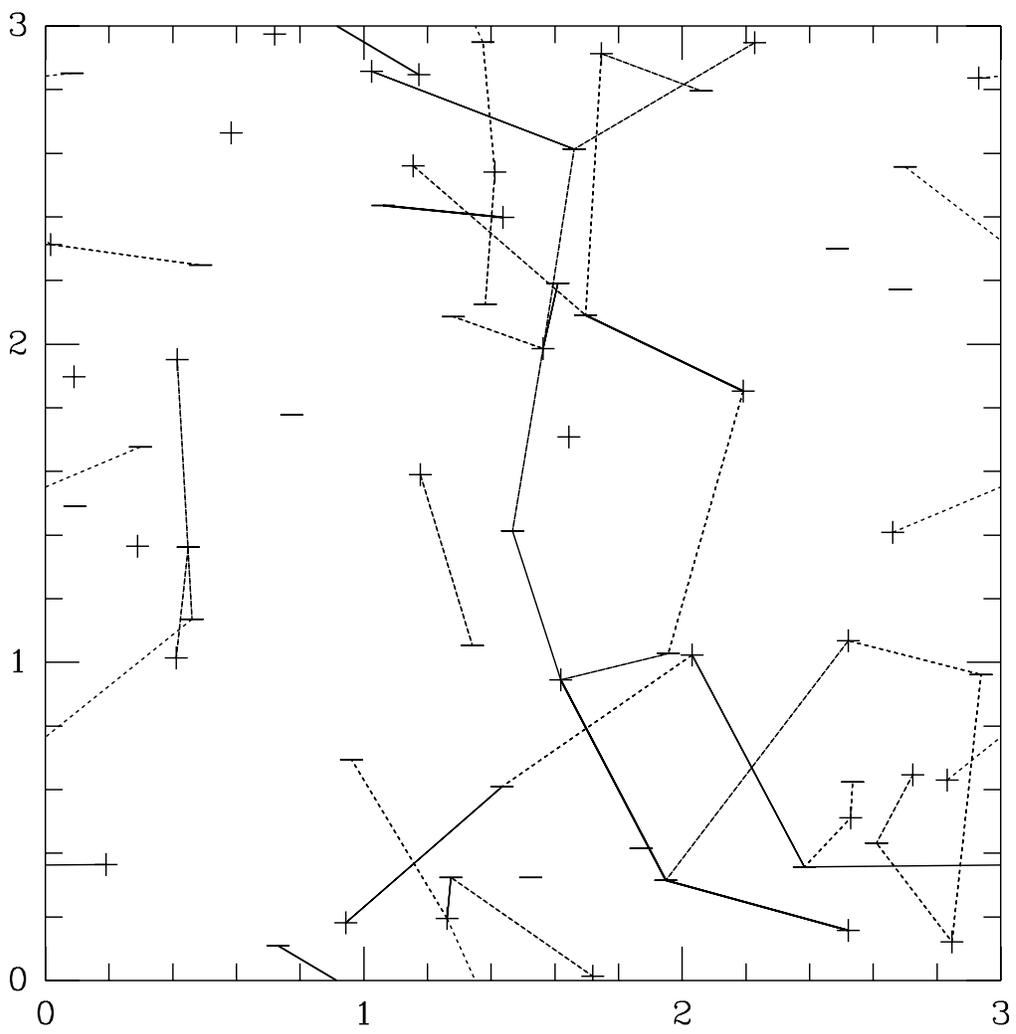
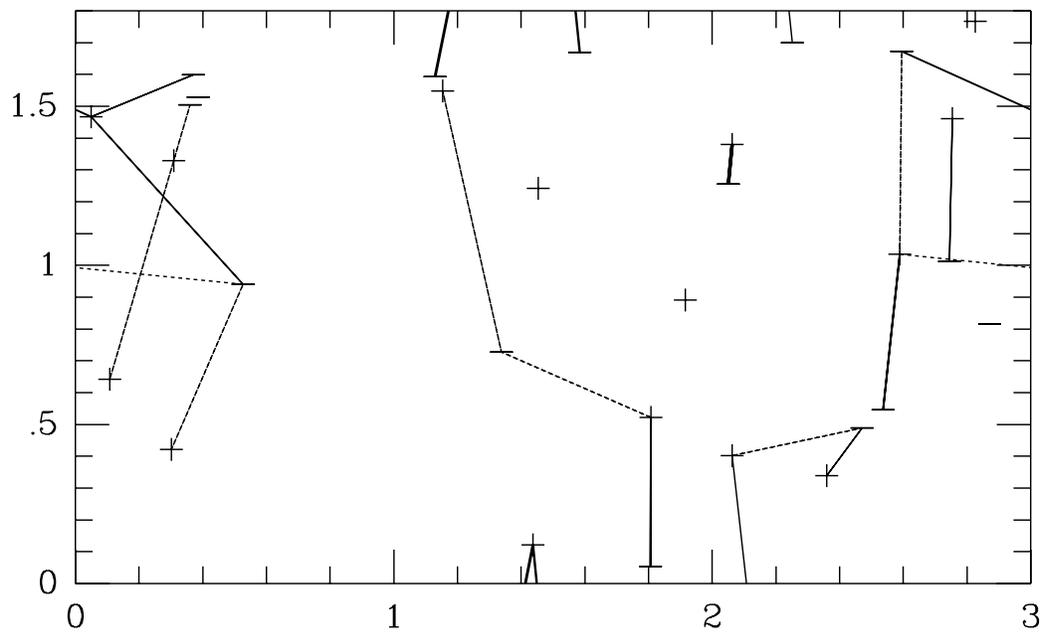
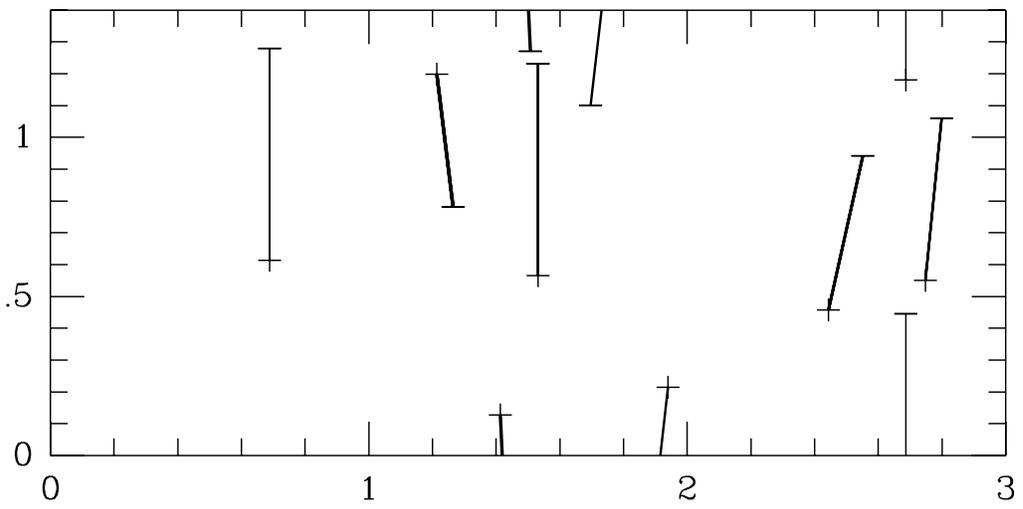
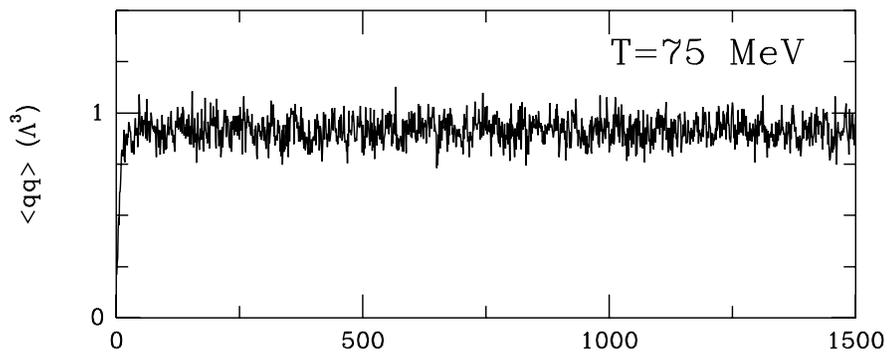
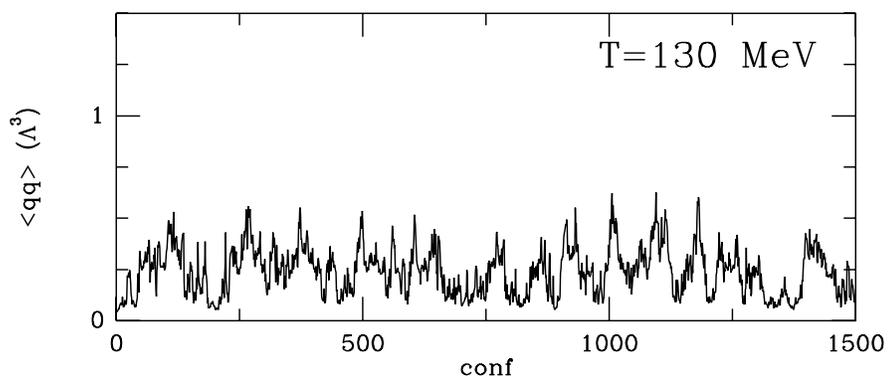
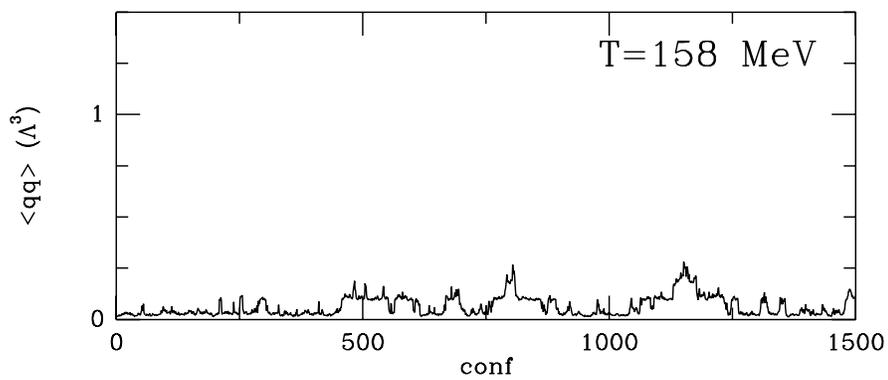
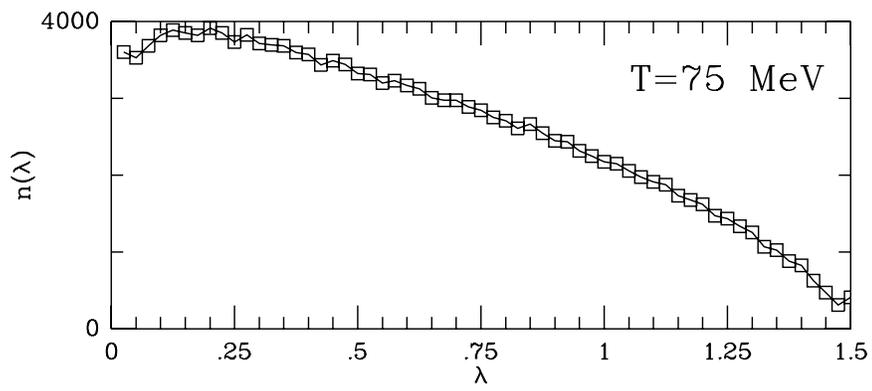
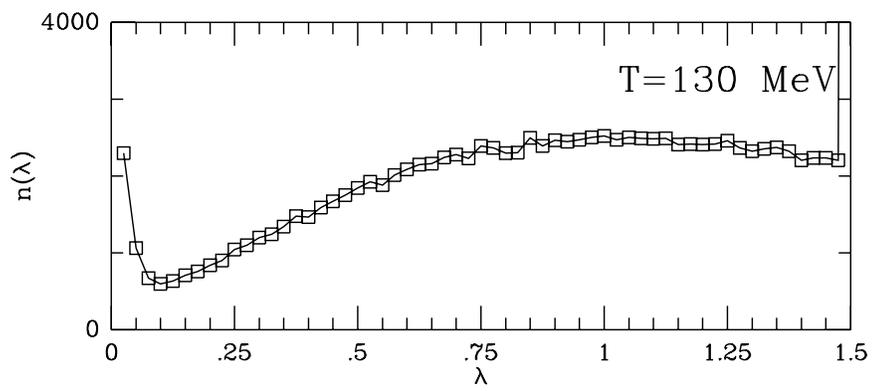
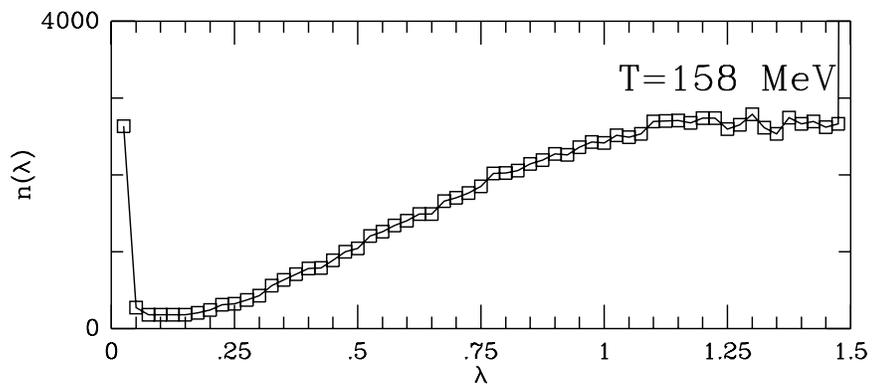
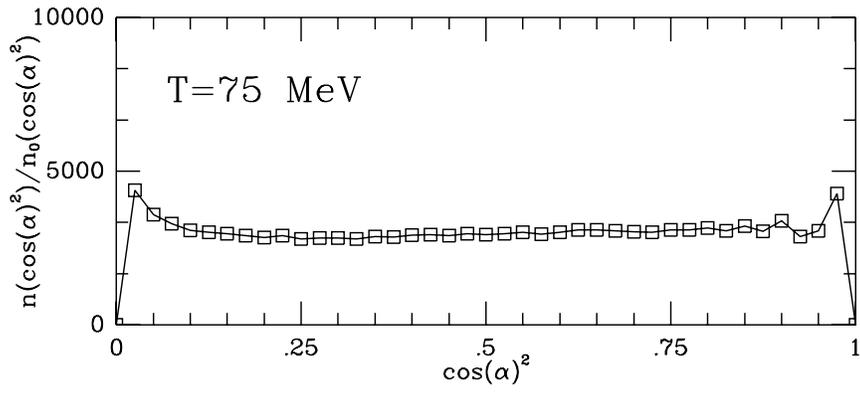
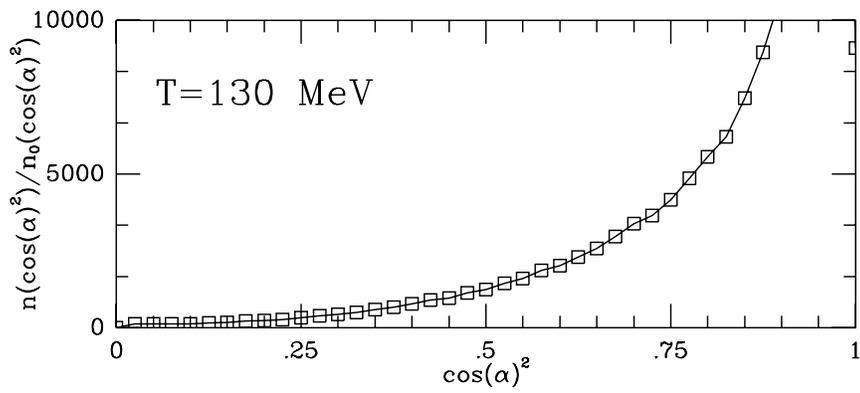
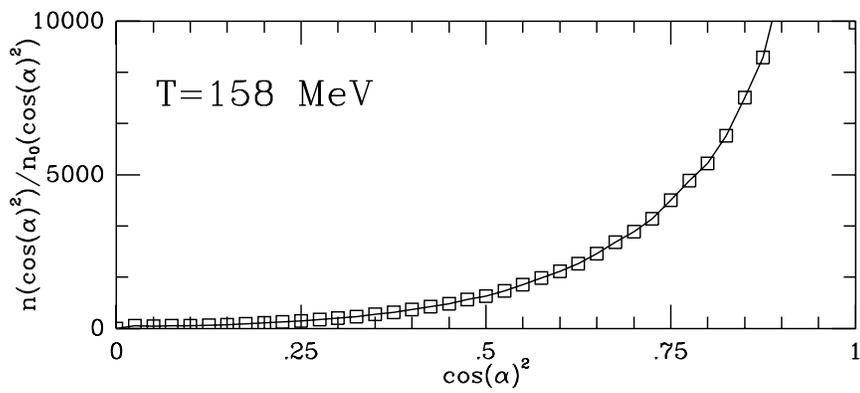
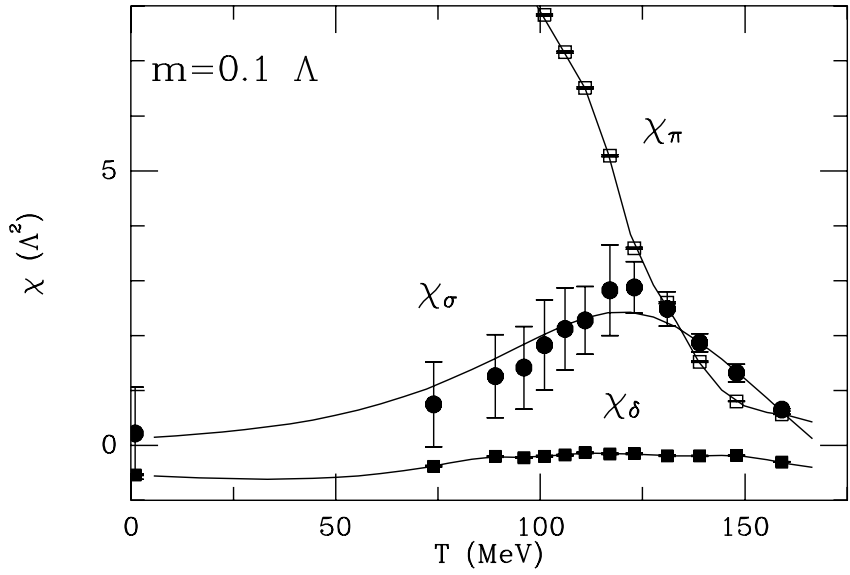
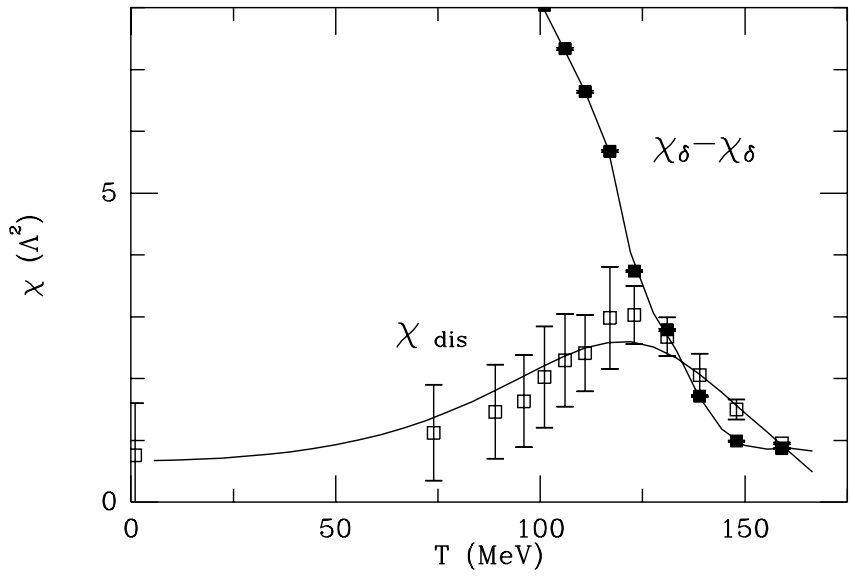
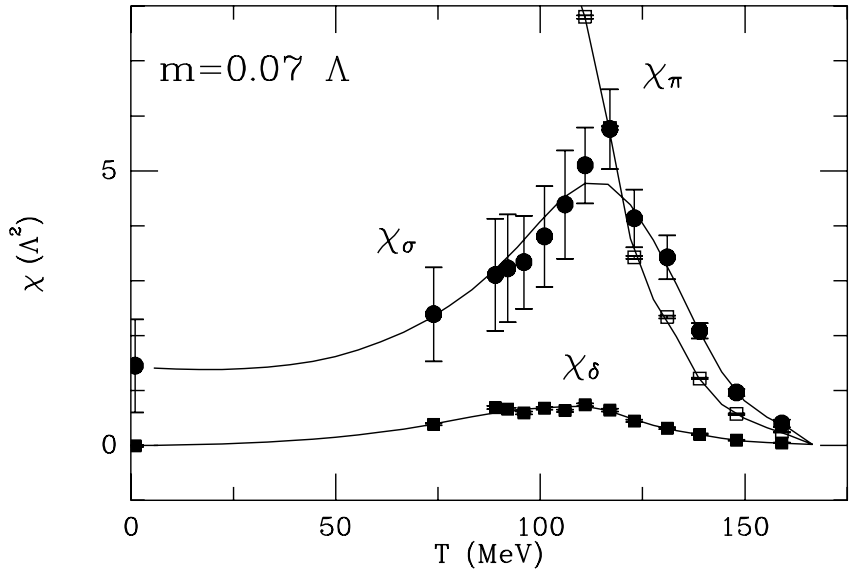
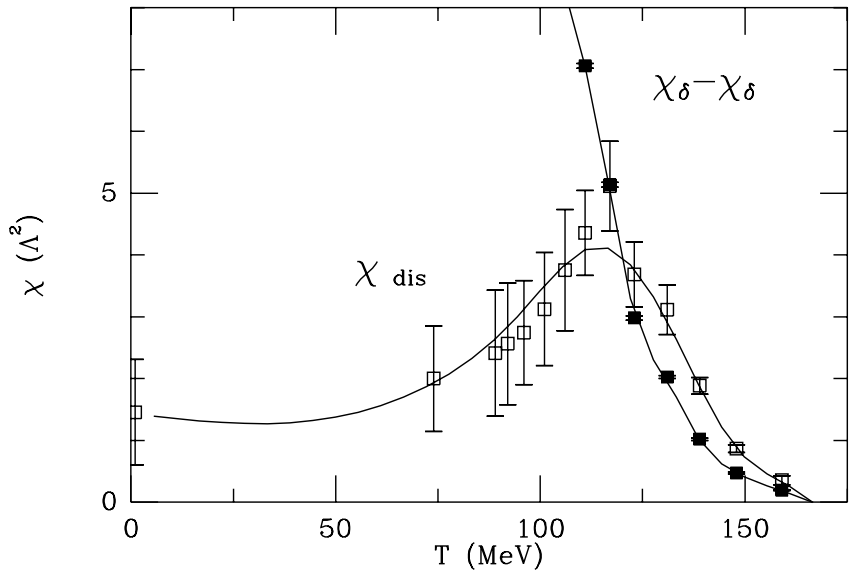
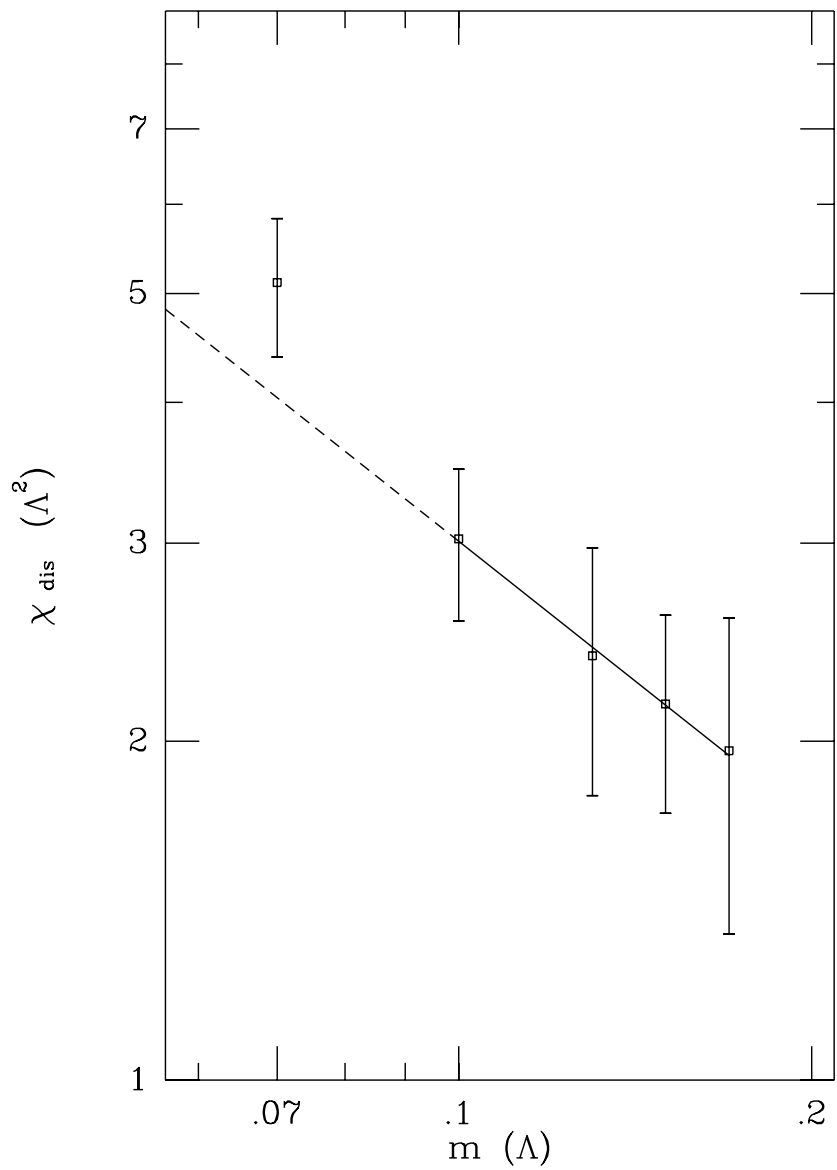
\end{document}